\documentclass[draft]{agujournal2019}
\usepackage{url} 
\usepackage{lineno}
\usepackage{placeins}
\usepackage{soul}
\usepackage{makecell}
\usepackage{amsmath} 
\usepackage{caption}
\usepackage{algorithmicx}
\usepackage{booktabs} 
\usepackage[inline]{trackchanges} 
\usepackage{natbib}
\usepackage{hyperref}
\usepackage{longtable}
\usepackage{graphicx}
\usepackage{tikz}
\usetikzlibrary{arrows.meta,positioning,shapes.geometric,fit,calc}
\usepackage{algorithm}
\usepackage{algpseudocode}
\usepackage{float}
\usepackage{lmodern}
\usepackage{enumitem} 
\usepackage{amssymb}
\usepackage{bm}

\draftfalse

\journalname{JGR: Machine Learning and Computation}

\begin{document}

\title{Forecasting Continuum Intensity for Solar Active Region Emergence Prediction using Transformers}

\authors{Jonas Tirona\affil{1}, Sarang Patil\affil{1}, Spiridon Kasapis\affil{2}, Eren Dogan\affil{1}, John Stefan\affil{1}, Irina N. Kitiashvili\affil{3}, 
Alexander G. Kosovichev\affil{1,3}, Mengjia Xu\affil{1}} 

\affiliation{1}{New Jersey Institute of Technology, Newark, NJ, USA}
\affiliation{2}{Princeton University, Princeton, NJ, USA}
\affiliation{3}{NASA Ames Research Center, Moffett Field, CA, USA}


\begin{abstract}
Early and accurate prediction of solar active region (AR) emergence is crucial for space weather forecasting. Building on established Long Short-Term Memory (LSTM) based approaches for forecasting the continuum intensity decrease associated with AR emergence, this work expands the modeling with new architectures and targets. We investigate a sliding-window Transformer architecture to forecast continuum intensity evolution up to 12 hours ahead using data from 46 ARs observed by SDO/HMI. We conduct a systematic ablation study to evaluate two key components: (1) the inclusion of a temporal 1D convolutional (Conv1D) front-end and (2) a novel `Early Detection' architecture featuring attention biases and a timing-aware loss function. Our best-performing model, combining the Early Detection architecture without the Conv1D layer, achieved a Root Mean Square Error (RMSE) of 0.1189 (representing a 10.6\% improvement over the LSTM baseline) and an average advance warning time of 4.73 hours (timing difference of -4.73h), even under a stricter emergence criterion than previous studies. While the Transformer demonstrates superior aggregate timing and accuracy, we note that this high-sensitivity detection comes with increased variance compared to smoother baseline models. However, this volatility is a necessary trade-off for operational warning systems: the model's ability to detect micro-changes in precursor signals enables significantly earlier detection, outweighing the cost of increased noise. Our results demonstrate that Transformer architectures modified with early detection biases, when used without temporal smoothing layers, provide a high-sensitivity alternative for forecasting AR emergence that prioritizes advance warning over statistical smoothness.
\end{abstract}

\section*{Plain Language Summary}
Solar active regions are areas of intense magnetic flux on the Sun that drive space weather events, such as solar flares, which can disrupt technologies on Earth. Forecasting the emergence of these regions is challenging because the warning signs, subtle variations in the Sun's acoustic and magnetic patterns, are faint and difficult to detect before the region becomes visible. In this study, we applied a Transformer machine learning architecture, designed to recognize complex sequential patterns, to analyze these precursor signals to predict the emergence of solar active regions. We compared standard models against a specialized "Early Detection" architecture. We found that standard architectures often prioritize statistical stability, and the addition of convolutional layers effectively smooth out the faint data spikes that signal the start of an event, causing inaccuracies in timing prediction. By modifying the model to be more sensitive to these initial fluctuations, our approach successfully predicted the emergence of active regions approximately 5 hours in advance. This demonstrates that for operational space weather forecasting, models must be designed to prioritize early detection of faint signals rather than prioritizing the smoothness of the prediction.

%
%

\section{Introduction} 
\label{sec:Introduction}
Solar active regions (ARs) manifest from the Sun's magnetic field emerging from the convection zone to the photosphere \citep[e.g.,][]{van2015evolution}, from where they drive events like space weather conditions by producing space weather events such as solar flares \citep[e.g.,][]{schrijver2009driving}, coronal mass ejections \citep[CMEs; e.g.,][]{howard1985coronal} and Solar Energetic Particles \citep[SEPs; e.g.,][]{whitman2023review}. These events, if sufficiently intense, can inject energetic particles in Earth's magnetosphere and affect technological infrastructure \citep[e.g.][]{horne2013space, kasapis2023turning,pak2025species}. Therefore, understanding when ARs emerge on the photosphere and produce flares, CMEs, and SEPs is crucial for predicting such geoeffective space-weather events and mitigating their effects on space-based infrastructure.

The emergence process of ARs involves rising magnetic flux through the solar interior, producing observable signatures in helioseismic data hours to days before becoming visible as sunspots \citep{kosovichev2001time, ilonidis2011detection, stefan2023exploring}. A number of helioseismic studies have confirmed that variations in the travel times of interior-penetrating acoustic waves can signal the upcoming flux emergence that later will form in ARs, before it can be observed visually on the photosphere \citep[e.g.,][]{birch2012helioseismology, leka2013helioseismology, barnes2014helioseismology, stefan2023exploring}. These travel time variations are extracted through Fourier and cross-correlation analyses of the surface velocity field, and it has been recently shown that direct use of the velocity field---without the use of computationally-expensive helioseismic inversion techniques---can also be effective in retrieving information about subsurface emerging magnetic flux \citep{kasapis2025prediction}, \citep{keegan2025data}. Recently, advances in machine learning and helioseismology have shown promise for predicting AR emergence using time-series analysis of solar oscillations observed on the solar surface \citep{kosovichev2025structure}. 

An initial architecture for studying such temporal sequence modeling was the Long Short-Term Memory (LSTM) network \citep{kasapis2023predicting}, originally introduced by \citet{hochreiter1997long}. Designed to mitigate the vanishing gradient problem inherent in standard Recurrent Neural Networks (RNNs), LSTMs utilize a gating mechanism to regulate the flow of information, making them particularly effective for learning order-dependence in time-series data. \citet{kasapis2023predicting,kasapis2025prediction} demonstrated that LSTMs could successfully forecast the continuum intensity associated with AR emergence by analyzing acoustic power and magnetic flux variations derived from the Helioseismic and Magnetic Imager (HMI) instrument data \citep{scherrer2012helioseismic, hoeksema2014helioseismic} onboard the Solar Dynamics Observatory (SDO) \citep{pesnell2012solar}, achieving predictions 10-29 hours in advance for several test cases. However, LSTMs are not immune to the limitations of recurrent architectures. Their sequential processing nature requires information to traverse every intermediate time step, which can cause the backpropagated error gradients—the signals used to update model weights—to diminish over extended sequences. This attenuation effectively prevents the model from learning relationships between distant time points.

To address these limitations and capture complex temporal dynamics more effectively, Transformer architectures~\citep{vaswani2017attention} have revolutionized sequence modeling across multiple domains through their self-attention mechanisms, which enable direct modeling of relationships between any two positions in a sequence regardless of their temporal distance. In time-series forecasting applications, Transformers have demonstrated superior performance over recurrent models by effectively capturing both local patterns and global dependencies \citep[e.g.,][]{zhou2021informer, wu2021autoformer}. For solar physics applications, the ability to simultaneously analyze multiple temporal scales may prove particularly valuable given the multi-scale nature of the Sun's magnetic activity.

This work extends the developments of predicting AR emergence by investigating the optimal architecture for a sliding-window Transformer. We present the first application of Transformer architectures to predict continuum intensity evolution driven by active region emergence. We conduct a systematic ablation study to evaluate two key components of our proposed forecasting framework: (1) the inclusion of a temporal 1D convolutional (Conv1D) front-end, and (2) a novel `Early Detection' architecture featuring specialized attention biases and a timing-aware loss function, alongside a detailed analysis of model behavior during the emergence phase. This paper is organized as follows: Section~\ref{sec:method} describes the dataset, preprocessing procedures, and the architectures of our experimental setup alongside the baseline LSTM model, detailing the loss functions and optimization strategies employed. Section~\ref{sec:results} presents quantitative performance metrics, hyperparameter search results, and detailed evaluation within emergence windows, supported by representative diagnostic plots. Section~\ref{sec:Discussion} interprets our findings, examining the trade-offs between statistical accuracy and physical plausibility of forecasts. Finally, Section~\ref{sec:conclusion} summarizes our key contributions and outlines directions for future research.

\section{Methodology} \label{sec:method}

\subsection{Dataset and Preprocessing}

Our analysis utilizes SolARED \citep{kasapis2025}, a machine learning-ready dataset, comprising 50 active regions observed with the SDO/HMI instrument. Each region, covering an area of $30.66^{\circ} \times 30.66^{\circ}$, was remapped onto the heliographic coordinates using the azimuthal equidistant (Postel's) projection centered its midpoint, and tracked using the Carrington rotation rate for a continuous 10-day period encompassing the phases surrounding emergence. From the original 50 regions, 4 were excluded due to data gaps and quality issues. The remaining 46 ARs were partitioned into a training/validation set of 41 regions and a held-out test set of 5 regions. While the source dataset contains 2D spatial maps ($512\times512$ pixels), we extract 1D time series for modeling by averaging the pixel values within each tile. The acoustic power maps ($P_a$) in the dataset are calculated from Doppler velocity measurements ($v_D$). The utilized dataset also includes line-of-sight magnetic field ($B_{los}$), and continuum intensity ($I_c$) maps. Each tracked region within each emerging AR is partitioned into a 9$\times$9 grid. Following \citet{kasapis2023predicting}, we primarily utilize the central row (9 tiles) of the tracked grid for model training to balance the distribution of quiet and emerging samples. However, since the model processes row-wise sequences independently, it can be applied to any row within the AR grid during inference. To correct for solar center-to-limb variations, the SolARED dataset normalizes the data relative to the top and bottom rows, effectively enhancing magnetic activity signals by subtracting systematic variations.

We applied normalization to the input dataset to ensure consistent scaling across different emergence events as
\begin{equation} \label{eq:normalization}
X' = \frac{X - X_{min}}{X_{max} - X_{min}},
\end{equation}
where $X$ represents the data (acoustic power bands and magnetic field), while $X_{min}$ and $X_{max}$ are global extremes computed over all tiles and observation periods. This normalization to [0, 1] ensures equitable feature contribution. Time-series sequences use a sliding window with input length $L_{in} = 110$ hours and prediction horizon $L_{out} = 12$ hours. For full details on geometric correction and tiling, see \cite{kasapis2025}.

\subsection{Transformer Architectures and Training}

To isolate the specific contributions of our architectural proposals, we employ a factorial experimental design \citep{fisher1935design}. This approach evaluates the interaction between two independent binary factors: (1) the inclusion of a convolutional front-end (Conv1D: True/False), and (2) the attention mechanism type (Standard vs. Early Detection). This yields four distinct model configurations, allowing us to determine if these components offer additive gains or act as functional replacements.

\begin{figure}[H]
\centering
\includegraphics[width=.86\textwidth]{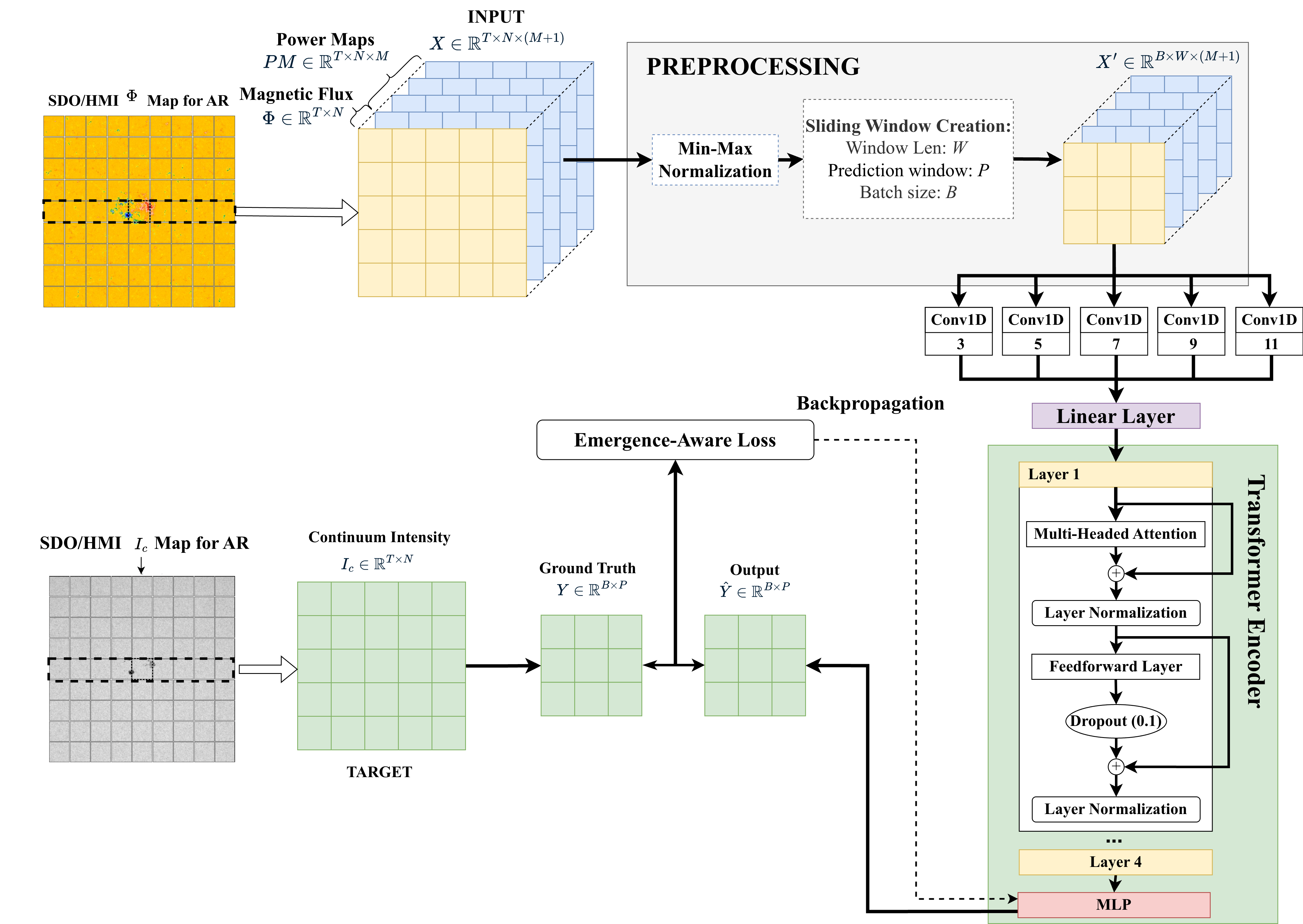}
\caption{Schematic overview of the end-to-end machine learning pipeline for continuum intensity decrease detection during the emergence of ARs. The system extracts central tiles from the tracked SDO/HMI magnetic flux ($\Phi$) map cut-outs, combining them with acoustic power maps to form feature tensor $X$. Sliding windows of length $W = 110$ are extracted with $P = 12$ prediction targets. An encoder-based Transformer predicts continuum intensity evolution $\hat{Y}$. Training utilizes an emergence-aware loss function optimizing MSE, early detection rewarding, and derivative loss.}
\label{fig:pipeline}
\end{figure}

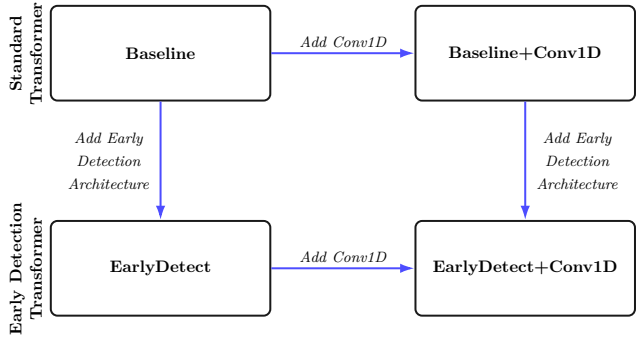
\begin{figure}[h]
\centering
\resizebox{0.6\textwidth}{!}{
    \begin{tikzpicture}[
      node distance=2.5cm and 3cm,
      exp_box/.style={rectangle, rounded corners, draw=black!90, very thick, align=center, inner sep=3mm, text width=4cm, minimum height=2cm},
      label_style/.style={font=\scshape\bfseries, align=center, text width=3.5cm},
      arrow_style/.style={-{Latex[length=3mm,width=2mm]}, very thick, draw=blue!70, font=\itshape\small, align=center}
    ]

    \node[exp_box] (expA) {\textbf{Baseline}};
    \node[exp_box, right=of expA] (expB) {\textbf{Baseline+Conv1D}};
    \node[exp_box, below=of expA] (expC) {\textbf{EarlyDetect}};
    \node[exp_box, right=of expC] (expD) {\textbf{EarlyDetect+Conv1D}};

    \node[label_style, left=0.5cm of expA, rotate=90, anchor=center] (row1) {Standard Transformer};
    \node[label_style, left=0.5cm of expC, rotate=90, anchor=center] (row2) {Early Detection Transformer};

    \draw[arrow_style] (expA.east) -- (expB.west) node[midway, above] {Add Conv1D};
    \draw[arrow_style] (expC.east) -- (expD.west) node[midway, above] {Add Conv1D};
    \draw[arrow_style] (expA.south) -- (expC.north) node[midway, left, xshift=-1mm] {Add Early \\ Detection \\ Architecture};
    \draw[arrow_style] (expB.south) -- (expD.north) node[midway, right, xshift=1mm] {Add Early \\ Detection \\ Architecture};

    \end{tikzpicture}%
}
\caption{
The factorial ablation study design. We compare two main architectures (Standard Transformer vs. Early Detection Transformer) and orthogonally evaluate the impact of a temporal 1D convolutional (Conv1D) front-end, resulting in four distinct model configurations: \texttt{Baseline}, \texttt{Baseline+Conv1D}, \texttt{EarlyDetect}, and \texttt{EarlyDetect+Conv1D}.
}
\label{fig:model_overview}
\end{figure}

We investigate a sliding-window Transformer and conduct a systematic ablation study to evaluate two key factors: the inclusion of a temporal 1D Convolutional (Conv1D) front-end, and the use of a specialized `Early Detection' architecture. The complete end-to-end data processing and forecasting pipeline is illustrated in Figure \ref{fig:pipeline}. This results in four distinct model configurations. Figure \ref{fig:model_overview} provides a visual summary of this experimental setup. The input to all model configurations consists of normalized time-series features $\mathbf{X} \in \mathbb{R}^{B \times L \times d_{in}}$, where $B$ denotes batch size, $L = 110$ hours represents the input sequence length, and $d_{in} = 5$ corresponds to four acoustic power maps at four frequency bands (2–3, 3–4, 4–5, and 5–6 mHz), and line-of-sight magnetograms. These features capture both the subsurface acoustic signatures preceding emergence and the evolving surface magnetic field configuration.

The baseline architecture (\texttt{Baseline} and \texttt{Baseline+Conv1D}) applies standard Transformer encoder layers to fixed-length temporal segments. The model processes input sequences $\mathbf{X} \in \mathbb{R}^{L_{in} \times d_{in}}$, where $d_{in} = 5$ denotes the number of input channels per time step. Input embedding is performed via linear projection:
\begin{equation}
\mathbf{H}^{(0)} = \mathbf{X}\mathbf{W}_{emb} + \mathbf{P},
\end{equation}
where $\mathbf{W}_{emb} \in \mathbb{R}^{d_{in} \times d_{model}}$ is the embedding matrix, and $\mathbf{P} \in \mathbb{R}^{L_{in} \times d_{model}}$ is a learnable positional embedding. The Transformer encoder applies $L$ layers of multi-head self-attention followed by feed-forward processing:
\begin{align}
\mathbf{A}^{(l)} &= \text{MultiHead}(\mathbf{H}^{(l-1)}),\\
\mathbf{H}^{(l)} &= \text{LayerNorm}(\mathbf{H}^{(l-1)} + \mathbf{A}^{(l)}), \\
\mathbf{F}^{(l)} &= \text{FFN}(\mathbf{H}^{(l)}), \\
\mathbf{H}^{(l)} &= \text{LayerNorm}(\mathbf{H}^{(l)} + \mathbf{F}^{(l)}),
\end{align}
where $\mathbf{H}^{(l)}$ denotes the output representation of the $l$-th encoder layer. $\text{MultiHead}(\cdot)$ represents the multi-head self-attention mechanism, which computes temporal dependencies across the sequence. $\text{FFN}(\cdot)$ is the position-wise feed-forward network consisting of two linear transformations with a non-linear activation, and $\text{LayerNorm}(\cdot)$ denotes layer normalization. The terms $\mathbf{A}^{(l)}$ and $\mathbf{F}^{(l)}$ represent the intermediate outputs of the attention and feed-forward sub-layers, respectively.

The \texttt{MultiHead} attention includes a learnable relative positional bias, as described by \citet{shaw2018self}. Finally, instead of standard global pooling, this model uses a weighted pooling mechanism. An auxiliary \texttt{emergence\_detector} head assigns a salience score to each time step, and the final sequence representation $\mathbf{h}_{pooled}$ is computed as the weighted sum of all time steps. This pooled vector is passed to a final \texttt{prediction\_head} to generate the 12-hour ahead forecast $\hat{\mathbf{y}}$. The baseline models (\texttt{Baseline} and \texttt{Baseline+Conv1D}) are trained with an \texttt{emergence\_aware\_loss} designed to optimize for accuracy and temporal smoothness:
\begin{equation}
\mathcal{L}_{Baseline} = \mathcal{L}_{MSE} + \lambda_1 \mathcal{L}_{deriv} + \lambda_2 \mathcal{L}_{temporal},
\label{eq:baseline_loss}
\end{equation}
where
\begin{align}
\mathcal{L}_{MSE} &= \frac{1}{N}\sum_{i=1}^{N}(\hat{y}_i - y_i)^2 \\
\mathcal{L}_{deriv} &= \frac{1}{N-1}\sum_{i=1}^{N-1}((\hat{y}_{i+1} - \hat{y}_i) - (y_{i+1} - y_i))^2 \\
\mathcal{L}_{temporal} &= \frac{1}{N-2}\sum_{i=1}^{N-2}|(\hat{y}_{i+2} - \hat{y}_{i+1}) - (\hat{y}_{i+1} - \hat{y}_i)|
\end{align}
Here, $\mathcal{L}_{deriv}$ penalizes inaccurate derivatives, while $\mathcal{L}_{temporal}$ encourages smooth predictions. Next, the Early Detection model (\texttt{EarlyDetect} and \texttt{EarlyDetect+Conv1D}) is a specialized architecture designed to explicitly favor earlier predictions. It enhances the baseline in two key ways:
\begin{enumerate}[label=(\arabic*)]
    \item \textbf{Enhanced Attention with Exponential Bias:} The \texttt{MultiHeadTemporalAttention} mechanism is modified to prioritize early temporal signals. We add a static exponential decay bias directly to the attention scores to systematically focus the model on the beginning of the sequence:
    \begin{equation}
        A_{i,j} = \frac{Q_i K_j^T}{\sqrt{d_k}} + \text{Bias}_{j}
    \end{equation}
    where $\text{Bias}_{j} = w_{early} \cdot \exp\left(\frac{-j}{L \cdot 0.3}\right)$. Here, $j$ represents the time step index, $L$ is the sequence length, and $w_{early}$ is a learnable weight. The scaling factor 0.3 controls the decay rate, effectively concentrating the attention bias within the first 30\% of the temporal window. This forces the mechanism to retain high sensitivity to earlier time steps where precursor signals reside.
    \item \textbf{Focused Pooling:} The standard weighted pooling is replaced with an early-focused pooling mechanism function. This function combines the auxiliary emergence scores with a fixed, non-learnable exponential decay weight, more aggressively weighting the start of the sequence when generating the pooled representation $\mathbf{h}_{pooled}$.
\end{enumerate}

The Early Detection models are trained with a composite objective function ($\mathcal{L}_{Early}$) that adds timing-specific penalties to the MSE baseline. The weights $\lambda_{timing}$ and $\lambda_{early}$ are hyperparameters optimized via grid search (evaluating $\lambda_{timing} \in \{0.2, 0.3, 0.4\}$ and $\lambda_{early} \in \{0.1, 0.2, 0.3\}$):
\begin{equation}
\mathcal{L}_{Early} = \mathcal{L}_{MSE} + \lambda_{timing} \mathcal{L}_{timing\_penalty} + \lambda_{early} \mathcal{L}_{early\_signal}
\label{eq:early_loss}
\end{equation}
where the specific components are defined as:
\begin{align}
\mathcal{L}_{timing\_penalty} &= \text{mean}(\text{ReLU}(\hat{\mathbf{y}} - \mathbf{y})) \\
\mathcal{L}_{early\_signal} &= \text{mean}(\text{ReLU}(\nabla\mathbf{y} - \nabla\hat{\mathbf{y}}))
\end{align}


The \textit{timing penalty} applies an asymmetric cost that strictly penalizes ``late'' predictions (where $\hat{y} > y$ during an intensity dip) while ignoring early predictions. This forces the optimization landscape to favor premature dips over delayed reactions. The \textit{early signal} term regularizes the derivative space, penalizing instances where the predicted gradient is shallower than the observed emergence gradient. Orthogonal to this core architecture, we evaluate a \texttt{TemporalConv1DBlocks} front-end to test the hypothesis that multi-scale 1D convolutions (kernels 3, 7, 15, 31) can extract short-term oscillatory precursors prior to global processing. Consequently, we compare models where this layer is enabled (\texttt{Baseline+Conv1D}, \texttt{EarlyDetect+Conv1D}) against those where the raw sequence is fed directly to the embedding layer (\texttt{Baseline}, \texttt{EarlyDetect}). These four model variants are summarized in Table \ref{tab:experimental_setup}.

\begin{table}[ht!]
\centering
\caption{The Factorial Experimental Setup.}
\label{tab:experimental_setup}
\scriptsize
\begin{tabular}{@{}lccc@{}}
\toprule
\textbf{Model Configuration} & \textbf{Core Architecture} & \textbf{Loss Function} & \textbf{Conv1D Enabled} \\
\midrule
\texttt{Baseline} & Baseline & Baseline & False \\
\texttt{Baseline+Conv1D} & Baseline & Baseline & True \\
\texttt{EarlyDetect} & Early Detection & Early Detection & False \\
\texttt{EarlyDetect+Conv1D} & Early Detection & Early Detection & True \\
\bottomrule
\end{tabular}
\end{table}

We employ two training methodologies: Baseline and Early Detection. A critical distinction from the LSTM baseline \citep{kasapis2023predicting}, which utilizes independent optimization per tile sequence, is our use of global parameter optimization. Our Transformer learns a single set of global weights across all active regions and tiles without shuffling. This constraint forces the model to learn spatially invariant representations, ensuring that attention patterns represent fundamental emergence signatures rather than tile-specific idiosyncrasies. The unified training framework is outlined in Algorithm~\ref{alg:transformer}.

\begin{algorithm}[h]
\caption{Unified Training Algorithm for Factorial Transformer Ablation Study}
\label{alg:transformer}
\footnotesize
\begin{algorithmic}[1]
\State \textbf{Input:} Training dataset $\mathcal{D}_{train}$, Model Configuration $M_{config}$
\State Initialize AdamW optimizer, OneCycleLR scheduler
\State
\For{each epoch}
    \For{each batch $(X, Y)$ in $\mathcal{D}_{train}$}
        \State Move batch to GPU
        \If{$M_{config} \in \{\texttt{Baseline+Conv1D}, \texttt{EarlyDetect+Conv1D}\}$} \Comment{Apply Conv1D}
            \State $X \gets \text{TemporalConv1DBlocks}(X)$
        \EndIf
        \State
        \If{$M_{config} \in \{\texttt{Baseline}, \texttt{Baseline+Conv1D}\}$} \Comment{Compute for Baseline arch}
            \State $\hat{Y} \gets \text{BaselineModel}(X)$
            \State $\mathcal{L} \gets \mathcal{L}_{Baseline}(\hat{Y}, Y)$ \Comment{Use loss from Eq. \ref{eq:baseline_loss}}
        \ElsIf{$M_{config} \in \{\texttt{EarlyDetect}, \texttt{EarlyDetect+Conv1D}\}$}
            \State $\hat{Y} \gets \text{EarlyDetectionModel}(X)$
            \State $\mathcal{L} \gets \mathcal{L}_{Early}(\hat{Y}, Y)$ \Comment{Use loss from Eq.~\ref{eq:early_loss}}
        \EndIf
        \State Zero gradients, backward, step optimizer, clip gradients, scheduler step
    \EndFor
    \State Perform validation and checkpoint the best model
\EndFor
\end{algorithmic}
\end{algorithm}

The unified training framework is outlined in Algorithm~\ref{alg:transformer}. The conditional logic (lines 7-17) explicitly implements the four branches of our ablation study. The model first selects whether to apply the \texttt{TemporalConv1DBlocks} preprocessing (lines 7-9). It then passes the features to either the \texttt{Baseline} or \texttt{EarlyDetect} architecture and computes the loss using the corresponding objective function ($\mathcal{L}_{Baseline}$ or $\mathcal{L}_{Early}$), as defined in Eqs. \ref{eq:baseline_loss} and \ref{eq:early_loss}.


\section{Results and Analysis} \label{sec:results}


\subsection{Implementation and Evaluation Protocol} \label{sec:eval}

All experiments were conducted using PyTorch 2.1.0 with CUDA 12.1 on NVIDIA A100 GPUs (40 GB memory) hosted on the NJIT High Performance Computing (HPC) facility. Training utilized mixed-precision arithmetic (FP16) via PyTorch's Automatic Mixed Precision (AMP) to accelerate computation while maintaining numerical stability. We used a batch size of 32 and the AdamW optimizer with OneCycleLR scheduling. The maximum learning rates were optimized individually for each model configuration (ranging from $5 \times 10^{-5}$ to $1 \times 10^{-3}$; see Appendix A for exact values), and gradient clipping with a maximum norm of 1.0 to stabilize training. Early stopping was applied based on validation loss with a patience of 50 epochs. We conducted a separate hyperparameter search for each of the four experimental configurations to find the optimal settings for each. The search space included embedding dimension ($d_{model}$), number of attention heads, number of encoder layers, dropout rate, and learning rate. For the Early Detection models, we also tuned the loss component weights, such as $\lambda_{early}$ and $\lambda_{timing}$. The final configurations used for our analysis represent the best-performing model from each of these distinct searches. 

The optimal architectural parameters resulting from this search are summarized in Table \ref{tab:model_architectures}. The complete hyperparameter search spaces and final detailed configurations for all models are provided in \ref{app:hyperparams} (Tables \ref{tab:hyperparams_baseline} through \ref{tab:best_config_lstm}).

\begin{table}[h]
\centering
\caption{Optimal architectural hyperparameters identified via grid search.}
\label{tab:model_architectures}
\scriptsize
\begin{tabular}{@{}lccccc@{}}
\toprule
\textbf{Model} & \textbf{Core Type} & \textbf{Conv1D} & \textbf{Encoder Layers} & \textbf{Attn. Heads} & \textbf{ $d_{model}$} \\
\midrule
LSTM & RNN & No & 3 & N/A & 64 (Hidden) \\
\texttt{Baseline} & Transformer & No & 3 & 4 & 128 \\
\texttt{Baseline+Conv1D} & Transformer & Yes & 5 & 4 & 256 \\
\texttt{EarlyDetect} & Transformer & No & 6 & 8 & 256 \\
\texttt{EarlyDetect+Conv1D} & Transformer & Yes & 4 & 4 & 512 \\
\bottomrule
\end{tabular}
\end{table}

Model performance was assessed on the held-out test ARs using a comprehensive, multi-model evaluation pipeline. This approach is designed to accurately separate the model's predictive timing from its physical accuracy by overlaying all models on the same plots.
\vspace{-0.07cm}
\begin{enumerate}[label=(\arabic*)]
    \item \textbf{Emergence Timing Error ($\Delta T$):} We calculate the temporal derivative of the normalized signal, $y_t$. The emergence onset time $t_{onset}$ is defined as the \textit{first} time step of a window where the smoothed derivative falls below a threshold $\delta = -0.01$ for a sustained duration of $k=4$ consecutive hours. This criterion is strictly enforced across all test regions to ensure robustness against false positives induced by transient solar noise. We note that this $k=4$ duration is more conservative than the $k>3$ hours threshold used by \citet{kasapis2023predicting}, ensuring that our reported advance-warning times represent sustained physical emergence. The timing error is calculated as $\Delta T = t_{onset}^{pred} - t_{onset}^{true}$. Consequently, a negative value for $\Delta T$ indicates an early (advance) prediction, while a positive value indicates a late forecast.

    \item \textbf{Emergence RMSE ($RMSE_{emerg}$):} Unlike overall RMSE, which is dominated by the quiet Sun baseline, this metric focuses exclusively on the dynamic emergence phase. We define the emergence window as a fixed 24-hour period starting from the detected onset: $[t_{onset}^{true}, t_{onset}^{true} + 24h]$.
    \begin{equation}
        RMSE_{emerg} = \sqrt{\frac{1}{24} \sum_{t=0}^{23} (y_{t_{onset}+t}^{denorm} - \hat{y}_{t_{onset}+t}^{denorm})^2}
    \end{equation}
    This metric is computed on denormalized (physical) values to ensure operational relevance.
\end{enumerate}

To ensure our metrics rigorously assess the model's ability to capture physical emergence rather than quiet-Sun background noise, we applied specific selection criteria to the test set evaluation. While the training set utilized the geometric center (Row 4), for the test Active Regions we selected the specific row intersecting the primary flux emergence site (varying between Rows 3, 4, and 5) to target the strongest physical signal. Furthermore, although the model outputs predictions for the full 9-tile width, we limit our quantitative reporting and visualization to the central 7 tiles (indices 1--7), excluding the quiescent leftmost and rightmost edges.

To facilitate a nuanced analysis of these targeted areas, we generated a comprehensive \textit{multi-tile diagnostic plot} for each test AR. Figure~\ref{fig:all_models_comparison_AR13183} shows the forecast delivered by all considered models (LSTM and all four Transformer configurations) across the seven central tiles of active region AR13183. Each tile features a five-panel display:
\vspace{-0.07cm}
\begin{itemize}
    \item \textbf{Intensity Forecast:} The main plot showing the denormalized, physical continuum intensity for the observed data and all five model predictions.
    \item \textbf{Observed Derivative:} The temporal derivative of the observed intensity, used for timing.
    \item \textbf{LSTM Derivative:} The temporal derivative of the LSTM prediction.
    \item \textbf{EarlyDetect Derivative:} The temporal derivative of the best-performing Transformer.
    \item \textbf{Absolute Error:} The absolute error of all five models over time.
\end{itemize}

This unified visualization allows for a direct, tile-by-tile qualitative comparison of model behavior, while the aggregated metrics (mean and standard deviation across all selected tiles for an AR) are exported to a separate table for quantitative analysis.

\subsection{Results}\label{sec:analysis}

We evaluated the best-performing model from each of the four experiments and the LSTM baseline on the held-out test dataset. Performance was measured using the metrics defined in our evaluation protocol (Section \ref{sec:eval}), including overall RMSE, emergence RMSE (RMSE computed only within the emergence window), and the emergence Timing Difference.
The results in Table \ref{tab:main_results} allow for a clear analysis of the two factors we tested. The optimal hyperparameters for the best-performing \texttt{EarlyDetect} model were identified as $\lambda_{timing}=0.4$ and $\lambda_{early}=0.3$, alongside a model architecture of 6 encoder layers, 8 attention heads, and an embedding dimension of 256. These optimized loss weights indicate that a stronger penalty on late predictions was necessary to force the temporal shift. The Early Detection architecture proved to be the most critical factor for success. This is best demonstrated by comparing the models without the Conv1D layer (\texttt{Baseline} vs. \texttt{EarlyDetect}). Upgrading from the \texttt{Baseline} to the \texttt{EarlyDetect} model yielded the most significant performance gain. It improved the Transformer's Overall RMSE (0.1239 $\to$ 0.1189), improved the Emergence RMSE (0.1586 $\to$ 0.1450), and critically, flipped the emergence timing from 8.27 hours \textit{late} to 4.73 hours \textit{early}—a total improvement of 13 hours, all while maintaining a similar parameter count (5.0M vs 5.8M). In contrast, when the Conv1D layer was present (\texttt{Baseline+Conv1D} vs. \texttt{EarlyDetect+Conv1D}), upgrading the architecture had an unclear impact: RMSEs were comparable (0.1303 $\to$ 0.1299), timing slightly worsened (+6.92h $\to$ +7.20h), and the parameter count increased substantially (5.6M $\to$ 19.1M).

\begin{table}[h]
\centering
\caption{Summary of the factorial ablation study. Bold values indicate the best performance for each metric. The \texttt{EarlyDetect} model (No Conv1D) achieved the best overall performance, including the lowest RMSE and the only negative (early) timing difference (in hours).}
\label{tab:main_results}
\footnotesize
\begin{tabular}{@{}lccccc@{}}
\toprule
\textbf{Metric} & \textbf{LSTM} & \textbf{Baseline} & \textbf{Baseline+Conv1D} & \textbf{EarlyDetect} & \textbf{EarlyDetect+Conv1D} \\
\midrule
Overall RMSE & 0.1330 & 0.1239 & 0.1303 & \textbf{0.1189} & 0.1299 \\
Timing Difference & +0.14h & +8.27h & +6.92h & \textbf{-4.73h} & +7.20h \\
Emergence RMSE & 0.1732 & 0.1586 & 0.1464 & \textbf{0.1450} & 0.1571 \\
Model Parameters & \textbf{184.6K} & 5.0M & 5.6M & 5.8M & 19.1M \\
\bottomrule
\end{tabular}
\end{table}

Conversely, the temporal Conv1D front-end was found to be detrimental. When added to the Baseline model (\texttt{Baseline} vs. \texttt{Baseline+Conv1D}), it worsened the Overall RMSE (0.1239 $\to$ 0.1303), though it modestly improved the emergence RMSE (0.1586 $\to$ 0.1464). The effect was similarly detrimental on the Early Detection model (\texttt{EarlyDetect} vs. \texttt{EarlyDetect+Conv1D}). Adding the Conv1D layer (\texttt{EarlyDetect+Conv1D}) hurt both the Transformer's overall RMSE (0.1189 $\to$ 0.1299) and emergence RMSE (0.1450 $\to$ 0.1571) and dramatically increased the model size (5.8M $\to$ 19.1M). In both cases, the Conv1D front-end added computational complexity and parameters while failing to improve—and in most cases, actively harming—model performance. Our study reveals a clear and decisive outcome. The specialized architectural biases and timing-aware loss function of the Early Detection models proved to be the most critical factors for generating operationally viable early predictions. A detailed comparison of the four experiments clarifies these findings:

\begin{itemize}[nosep]
\item \textit{Baseline Models (\texttt{Baseline} \& \texttt{Baseline+Conv1D})}: These models served as a crucial control. While their overall RMSEs were respectable (0.1239 and 0.1303, respectively), their inability to predict the emergence in advance (+8.27h and +6.92h timing) highlights the limitations of a standard Transformer architecture, even when trained with a derivative-aware loss.
\item \textit{Early Detection + Conv1D (\texttt{EarlyDetect+Conv1D})}: This configuration had the highest parameter count by a wide margin (19.1M) while showing degraded RMSE (0.1299) and the second-worst timing (+7.20h).
\item \textit{Early Detection - Conv1D (\texttt{EarlyDetect})}: This model yielded the strongest aggregate performance. By utilizing the specialized Early Detection architecture without the Conv1D layer, it achieved the best performance in every key metric: the lowest Overall RMSE (0.1189), the lowest Emergence RMSE (0.1450), and a highly efficient parameter count (5.8M). Most importantly, it was the \textit{only} configuration to achieve an advanced forecast, predicting the emergence 4.73 hours early.
\end{itemize}

To validate that the timing improvements were not driven by outliers or local tile variances, we analyzed the distribution of forecasting errors across the five test active regions (ARs 11698, 11726, 13165, 13179, 13183). For each AR, we calculated the mean timing difference between the predicted and observed onset of emergence ($\Delta T$) across the seven central tiles to generate a single representative metric per region. Table \ref{tab:forecasting_patterns} details the statistics of these AR-level aggregates ($N=5$), highlighting the median timing shift and the percentage of regions successfully forecasted in advance.

\begin{table}[h]
\centering
\caption{Prediction forecasting patterns. This table details the distribution of timing errors across the five test ARs. Bold values indicate the best performance for each metric. The best-performing model configuration (\texttt{EarlyDetect}) and its superior metrics (earliest mean/median timing and highest percentage of early forecasts) are shown in bold.}
\label{tab:forecasting_patterns}
\footnotesize
\begin{tabular}{@{}lcccccc@{}}
\toprule
\textbf{Model} & \textbf{Mean $\Delta$T} & \textbf{Median $\Delta$T} & \textbf{Std. Dev.} & \textbf{\% Early} & \textbf{\% Late} \\
& (hrs) & (hrs) & (hrs) & ($<0$h) & ($>0$h) \\
\midrule
LSTM & 0.36 & 0.67 & 14.76 & 40.0 & 60.0 \\
\texttt{Baseline} & 3.71 & 8.75 & 12.94 & 40.0 & 60.0 \\
\texttt{Baseline+Conv1D} & 6.92 & 11.00 & 11.98 & 40.0 & 60.0 \\
\textbf{EarlyDetect} & \textbf{-4.73} & \textbf{-9.40} & 14.64 & \textbf{60.0} & \textbf{40.0} \\
\texttt{EarlyDetect+Conv1D} & 7.20 & 8.00 & 13.26 & 40.0 & 60.0\\
\bottomrule
\end{tabular}
\end{table}

The data in Table \ref{tab:forecasting_patterns} reinforce the improved global detection capability of the EarlyDetect architecture, though it also highlights a trade-off between stability and sensitivity. While the baseline models and the Conv1D variants skew towards late predictions (median offsets ranging from +8.00h to +11.00h), the \texttt{EarlyDetect} model achieves a median timing difference of -9.40 hours. Furthermore, it is the only configuration that achieves a majority of early forecasts (60\%), compared to the 40\% rate observed in all other models. The standard deviation (14.64 hours) and high variance in per-AR metrics (e.g., AR 11726 RMSE $\pm 956$) reflect this sensitivity trade-off: the model's noise is a direct consequence of its ability to react to micro-changes in precursor signals. This volatility is operationally preferable to the false stability of over-smoothed models that miss early onset entirely. The \texttt{EarlyDetect} model systematically shifts the predictive distribution toward earlier detection, with the magnitude of lead time varying based on the strength and clarity of precursor signatures in each active region. Detailed performance metrics for the five test active regions is presented in \ref{app:per_ar_results} (Tables \ref{tab:ar11698_metrics}–\ref{tab:ar13183_metrics}). 

We also evaluated the performance of the models for each testing active region. For instance, Table \ref{tab:qualitative_metrics_AR13183} provides the mean metrics and standard deviations for  AR13183, averaged across the seven central tiles. These metrics are computed on the denormalized, physical data. Specifically, we reverse the min-max scaling (from the $[0, 1]$ range) to restore physical intensity units to enable comparison with observations. 

\begin{table}[h]
\centering
\caption{Mean qualitative metrics for AR13183. Metrics are averaged across the central tiles 38-44. Bold values indicate the best performance for each metric.}
\label{tab:qualitative_metrics_AR13183}
\footnotesize
\begin{tabular}{@{}llccccc@{}} 
\toprule
\textbf{Metric} & & \textbf{LSTM} & \textbf{Baseline} & \makecell{\textbf{Baseline}\\\textbf{+Conv1D}}
 & \textbf{EarlyDetect} & \makecell{\textbf{EarlyDetect}\\\textbf{+Conv1D}} \\ 
\midrule
Overall RMSE & Mean & 416.41 & 328.78 & 410.38 & \textbf{328.27} & 389.59 \\
             & (Std) & (268.03) & (165.90) & (222.10) & \textbf{(143.74)} & (175.88) \\
\addlinespace
Timing difference (h) & & +7.50 & +8.75 & +11.33 & \textbf{+4.75} & +8.00 \\
\addlinespace
Emergence RMSE & Mean & \textbf{516.80} & 667.83 & 807.40 & 516.93 & 706.81 \\
               & (Std) & (276.63) & \textbf{(44.04)} & (252.02) & (146.41) & (209.34) \\
\addlinespace
Model parameters & & \textbf{184.6K} & 5.0M & 5.6M & 5.8M & 19.1M \\
\bottomrule
\end{tabular}
\end{table}

\begin{figure}[H]
\centering
\includegraphics[width=\textwidth]{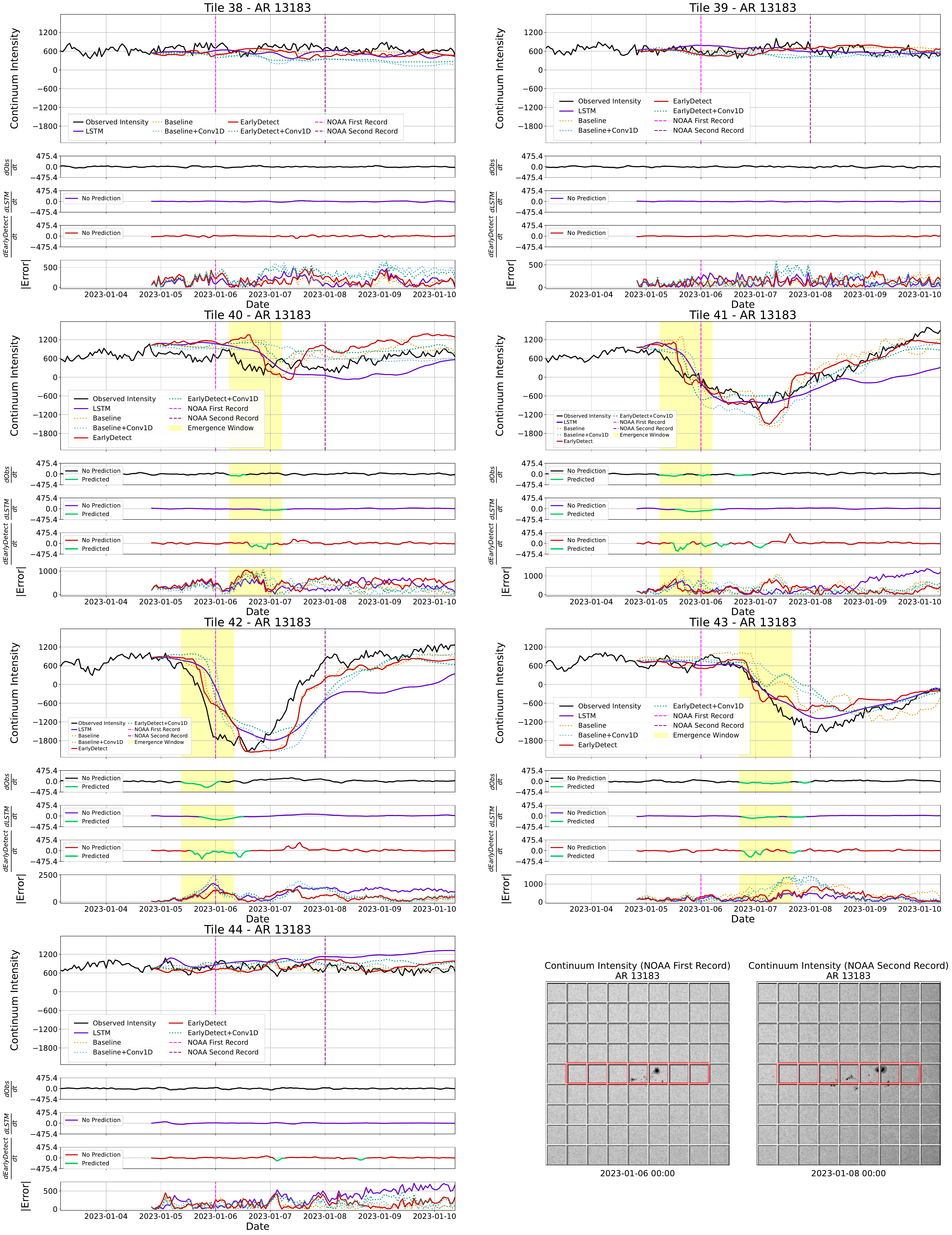}
\caption{Comparison of predicted and actual evolution of the continuum intensity of AR13183 for seven central tiles (indicated in the right bottom corner). Each tile includes five subplots: (1) Main intensity forecast with all models overlaid, (2) Observed derivative, (3) LSTM derivative, (4) \texttt{EarlyDetect} derivative, and (5) Absolute error for all models. The lower-right corner includes two 9x9 continuum maps showing the AR at the start and end of the plotted window. The presented results are denormalized to physical units and reflect variations relative to the quiet Sun, which is considered as background.}
\label{fig:all_models_comparison_AR13183}
\end{figure}

For this specific test AR, the \texttt{EarlyDetect} model (5.8M parameters) was most operationally useful, achieving the best predictive timing (+4.75 hours) and the lowest Overall RMSE (328.27). This case highlights the difference between statistical fit (Overall RMSE) and the capture of event dynamics (Timing/Emergence RMSE). The \texttt{Baseline} model, for example, had a similar Overall RMSE (328.78) but far worse timing (+8.75 hours) and Emergence RMSE (667.83). Visual inspection of the emergence profile from Figure~\ref{fig:all_models_comparison_AR13183} reveals that AR 13183 exhibits a notably smoother continuum intensity drop-off and a more gradual return to the mean compared to the sharper, more volatile transitions observed in other test regions (e.g., AR 11698). Despite this naturally smooth morphology, the Conv1D layer's smoothing effect remains detrimental: \texttt{EarlyDetect+Conv1D} (19.1M parameters) achieved worse timing (+8.00h), higher Overall RMSE (389.59), and higher Emergence RMSE (706.81) compared to \texttt{EarlyDetect}. This demonstrates that even for events with inherently smooth profiles, the Conv1D layer's temporal filtering degrades performance by averaging out critical precursor signals that the Early Detection attention mechanism is designed to capture.

While previous metrics represent spatial averages across central tiles, Table \ref{tab:emergence_summary_short} details tile-level alarm timing. This breakdown clarifies the divergence between LSTM stability and Transformer sensitivity. The LSTM tends to issue synchronous alarms, yielding low variance but late predictions. In contrast, EarlyDetect exhibits significant per-tile variance, driven by its sensitivity to local precursors. For instance, in AR11698, it triggers alarms as early as $-70$ hours on specific tiles, while the LSTM remains reactive. This volatility is inherent to high-sensitivity detection: the model prioritizes early warning over uniformity, accepting localized false positives for the critical benefit of earlier detection—an essential trade-off for operational forecasting.

\begin{table}[h]
\centering
\caption{Comparison of emergence alarms (in hours) for the best-performing Transformer(\texttt{EarlyDetect}) vs. the LSTM Baseline. Note: Negative values indicate early (advance) detection; positive values indicate late detection. "Quiet" indicates correct rejection. FP = False Positive; FN = False Negative.}
\label{tab:emergence_summary_short}
\footnotesize
\begin{tabular}{l|cccccccc}
\toprule
\textbf{AR (Tiles)} & \textbf{Model} & \textbf{T1} & \textbf{T2} & \textbf{T3} & \textbf{T4} & \textbf{T5} & \textbf{T6} & \textbf{T7} \\
\midrule
\textbf{11698} & LSTM & Quiet & FN & 8h & 61h & 12h & 11h & Quiet \\
(47--53) & \texttt{EarlyDetect} & Quiet & FN & -41h & 61h & -70h & 11h & FP \\
\midrule
\textbf{11726} & LSTM & Quiet & Quiet & 14h & -27h & 3h & -27h & -6h \\
(38--44) & \texttt{EarlyDetect} & Quiet & Quiet & 18h & -31h & 0h & -27h & -7h \\
\midrule
\textbf{13165} & LSTM & Quiet & Quiet & -33h & -12h & 5h & -43h & Quiet \\
(29--35) & \texttt{EarlyDetect} & Quiet & Quiet & -67h & 2h & 6h & -46h & FP \\
\midrule
\textbf{13179} & LSTM & Quiet & Quiet & -8h & 0h & 10h & FP & Quiet \\
(38--44) & \texttt{EarlyDetect} & Quiet & Quiet & 21h & 16h & 14h & Quiet & Quiet \\
\midrule
\textbf{13183} & LSTM & Quiet & Quiet & 14h & 7h & 8h & 1h & Quiet \\
(38--44) & \texttt{EarlyDetect} & Quiet & Quiet & 9h & 5h & 4h & 1h & FP \\
\bottomrule
\end{tabular}
\end{table}

Having established this context, we present a comprehensive diagnostic plot in Figure \ref{fig:all_models_comparison_AR13183} to visually confirm these findings. This unified visualization overlays five models across the seven central tiles, allowing for a direct, tile-by-tile qualitative comparison. The figure corroborates the metrics in Table \ref{tab:qualitative_metrics_AR13183}: the \texttt{EarlyDetect} (red) prediction aligns best with the observed drops, while the \texttt{Baseline+Conv1D} is visibly late. Supplementary plots for the remaining test ARs following this same format are provided in \ref{app:plots} (Figures \ref{fig:ar_11698}–\ref{fig:ar_13179}).

To evaluate the practical trade-offs of our four Transformers models, we profiled their computational and memory costs. We compared them against the baseline LSTM and measured: (1) Computational Complexity in floating-point operations (FLOPs), (2) peak GPU memory usage during a forward pass, and (3) inference latency (Forward Pass Time). The results are detailed in Table \ref{tab:complexity_results}.
\begin{table}[h]
\centering
\caption{Model complexity and efficiency metrics.}
\label{tab:complexity_results}
\footnotesize
\begin{tabular}{@{}lcccc@{}}
\toprule
\textbf{Model} & \textbf{\# Parameters} & \textbf{FLOPs} & \textbf{Peak Memory} & \textbf{Fwd. Pass Time} \\
&  & (G) & (MB) & (ms) \\
\midrule
LSTM & 184.6K & 0.69 & 55.5 & 2.79 \\
\texttt{Baseline} & 5.0M & 40.4 & 483.5 & 5.50 \\
\texttt{Baseline+Conv1D} & 5.6M & 44.8 & 538.6 & 6.40 \\
\texttt{EarlyDetect} & 5.8M & 46.8 & 661.4 & 7.14 \\
\texttt{EarlyDetect+Conv1D} & 19.1M & 153.1 & 1042.0 & 11.81 \\
\bottomrule
\end{tabular}
\end{table}

The analysis reveals several key findings:
\begin{itemize}
\item \textbf{Overall Cost:} All Transformer models are significantly more expensive than the LSTM baseline, requiring 8-18 times more memory and 60-220 times more FLOPs.
\item \textbf{Impact of Conv1D:} The Conv1D layer consistently adds overhead. Comparing the Baseline model to \texttt{Baseline+Conv1D}, the Conv1D layer adds 0.6M parameters, 4.4G FLOPs, 55MB of memory, and ~1ms of latency.
\item \textbf{Cost of Early Detection:} The Early Detection architecture (\texttt{EarlyDetect}) is slightly more costly than the simple \texttt{Baseline}, adding 0.8M parameters and ~1.6ms of latency, but is comparable in efficiency.
\item \textbf{The Worst Case (\texttt{EarlyDetect+Conv1D}):} The combination of the Early Detection architecture and the Conv1D layer results in a catastrophic increase in complexity. This model is 3.3 times larger than our best model (\texttt{EarlyDetect}), requires 3.3 times the FLOPs, and is the slowest to run.
\end{itemize}
This analysis strongly reinforces our earlier findings. The optimal model, \texttt{EarlyDetect}, provides  superior accuracy and timing (a -4.73h forecast) for only a minor computational cost compared to the \texttt{Baseline}. Conversely, \texttt{EarlyDetect+Conv1D} is the worst-performing model, as it is both highly inaccurate (a +7.20h lag) and by far the most computationally expensive.

\section{Discussion} \label{sec:Discussion}
The comparative ablation study yielded a clear trade-off between smoothness and sensitivity. While the LSTM baseline and Baseline Transformer models produce smoother, visually appealing forecasts, they consistently lag behind the actual emergence, rendering them operationally inadequate for advance warning. The specialized Early Detection Transformer (\texttt{EarlyDetect}) was the only model to achieve the primary operational objective: a true forecast with an average lead time of $-4.73$ hours. The analysis of the four experiments reveals a critical decoupling of statistical accuracy (RMSE) from predictive timing ($\Delta$T). The baseline Transformer models (\texttt{Baseline} \& \texttt{Baseline+Conv1D}) achieved respectable RMSE scores but produced forecasts that were operationally useless, lagging the actual emergence by $+6.92$ to $+8.27$ hours. Similarly, the baseline LSTM, while computationally efficient, achieved only synchronous detection ($+0.14$h) and failed to provide a true advance warning. This demonstrates that a standard model, when optimized with a generic or even a derivative-aware loss function ($\mathcal{L}_{Baseline}$), will learn to produce a statistically accurate ``best fit'' of the emergence event, which it achieves by placing the predicted dip after the event has already begun. The success of \texttt{EarlyDetect}, therefore, is almost entirely attributable to the specialized \texttt{early\_detection\_loss} ($\mathcal{L}_{Early}$) and model attention biases. These components successfully forced the model to abandon a ``perfect'' physical fit and instead prioritize the subtle, early-onset signatures, achieving the primary goal of a pre-emergence warning. Remarkably, this shift in priority did not compromise overall accuracy; \texttt{EarlyDetect} simultaneously achieved the lowest RMSE of all models (0.1189), demonstrating that high-sensitivity forecasting can yield superior statistical performance alongside operational utility.

Our results reveal a critical distinction between smooth, conservative predictions and operationally useful early warnings. While the LSTM baseline and Baseline Transformer models produce smoother, visually appealing forecasts, they consistently lag behind the actual emergence. For instance, in AR 13165, the \texttt{Baseline} model predicted the emergence $+10.5$ hours late (averaged across tiles), while \texttt{EarlyDetect} achieved a mean advance warning of $-26.25$ hours—a difference of over 36 hours. Similarly, in AR 11698, the LSTM baseline was on average $+23.0$ hours late compared to \texttt{EarlyDetect}'s mean $-9.75$ hour early detection. This pattern holds across the aggregate metrics: while the LSTM achieves synchronous detection (global mean $+0.14$h), the \texttt{EarlyDetect} Transformer systematically provides advance warnings (mean $-4.73$h) with a $60\%$ early detection rate versus the LSTM's $40\%$. The LSTM's per-tile optimization strategy, while producing visually stable fits, fundamentally fails to capture the subtle precursor signals that precede emergence. In contrast, the \texttt{EarlyDetect} model's attention mechanism is specifically tuned to detect these early signatures, resulting in higher variance but operationally critical lead times. This higher variance—evident in metrics like AR 11726's RMSE standard deviation of $\pm 956$—is a necessary trade-off for sensitivity: the model is ``noisy'' because it reacts to micro-changes in the acoustic power signatures, whereas the LSTM and Conv1D models are ``over-smoothed'' and miss the earliest onset signals entirely.

This difference in stability is further influenced by the training scope. The LSTM baseline benefited from a per-tile optimization strategy (fresh optimizer per sequence), allowing it to adapt closely to specific local dynamics. In contrast, our Transformer was constrained to a single global optimizer across all regions without shuffling. While this global approach likely contributed to the higher variance in specific edge cases (e.g., AR 11698) as the model could not overfit to local variations, the fact that the \texttt{EarlyDetect} model still achieved a 10.6\% improvement in RMSE suggests it has learned a more fundamental, generalized representation of emergence physics.

Conversely, the temporal 1D convolutional (Conv1D) front-end was found to be decisively detrimental. Our hypothesis was that the Conv1D layer would help by extracting local, temporal features from the input sequences. The results from the \texttt{EarlyDetect+Conv1D} model prove this hypothesis wrong. In the most extreme case (AR 11698), the addition of the Conv1D layer collapsed the mean forecast from a $-9.75$ hour advance warning to a $+26.00$ hour delay. The Conv1D layer not only caused a substantial increase in model parameters and computational cost (from 5.8M to 19.1M) but also worsened all key metrics, including a 13-hour degradation in timing compared to \texttt{EarlyDetect} ($-4.73$h $\to$ $+7.20$h). We interpret this as the Conv1D layer ``over-smoothing'' the input data, effectively averaging out the very faint precursor signals that the Early Detection attention mechanism was designed to find. The unique performance of AR 13183 supports this ``over-smoothing'' hypothesis by acting as the exception that proves the rule. As noted, AR 13183 is characterized by a naturally smooth drop-off and recovery profile. In this specific instance, the smoothing properties of the convolutional front-end likely aligned with the physical morphology of the event, acting as a beneficial filter against high-frequency noise. However, for the majority of active regions which are defined by rapid, sharp, and faint emergence signatures, this same smoothing effect washes out the critical precursor signals, resulting in the degraded performance seen in the aggregate metrics. 

The optimal model (\texttt{EarlyDetect}) performed best when fed the raw, unfiltered time series. This suggests that the self-attention mechanism is superior to both convolutional layers and the recurrent gates of the LSTM at isolating the specific acoustic power signatures (specifically in the 3-5 mHz range) that precede emergence, without requiring pre-filtering or manual band selection often necessary in prior works \citep[e.g.,][]{ilonidis2011detection, leka2013helioseismology}. Unlike the smoothed models, \texttt{EarlyDetect} exhibits a characteristic pre-emergence dip in its predictions (e.g., AR 13165 Tile 31; AR 11698 Tiles 49 \& 51), effectively ``flagging'' the region hours before the flux officially rises. This qualitative signal shape—a subtle but persistent downward trend preceding the main intensity drop—is the direct result of the model's attention biases and timing-aware loss function, which force the architecture to prioritize early-onset signatures over statistical fit. This demonstrates that for this forecasting task, a specialized architecture that directly optimizes for predictive timing is superior to a generic model. Furthermore, the failure of the 
\texttt{EarlyDetect+Conv1D} model suggests a conflict between the two architectural components. This model had the highest parameter count (19.1M), a degraded RMSE (0.1299), and poor timing (+7.20h). We hypothesize that the features extracted by the Early Detection attention mechanism are made redundant or, worse, are actively harmed by the Conv1D layer's smoothing, resulting in a bloated and inefficient model. 

Our evaluation yields a mean timing difference of $+0.14$ hours for the LSTM baseline, which differs from the lead times of $-10$ to $-29$ hours reported by \citet{kasapis2023predicting}. This discrepancy is attributable to the standardized evaluation protocol employed in this study. While we utilized the same per-active-region input window lengths (e.g., 72 hours for AR11726, 96 hours for others) to account for limb proximity, we enforced a uniform emergence criterion ($k=4$ hours, $\delta = -0.01$) and a fixed metric calculation across the entire test set. Unlike previous approaches that may have benefited from per-event threshold tuning, our standardized protocol prevents the exploitation of ``lucky'' early precursors. Under these rigorous conditions, the LSTM's aggregate behavior converges on the emergence onset (synchronous detection), despite achieving significant lead times in isolated cases (e.g., AR 13165). In contrast, the Transformer systematically shifts the predictive distribution toward earlier detection, reinforcing the necessity of the attention mechanism to isolate faint precursor signals consistently.

It is crucial to distinguish between the experimental lead times reported here and the constraints of a real-time operational environment. \citet{kasapis2023predicting} identified the operational latency inherent to this pipeline, primarily driven by the time required for Dopplergram sequence processing. However, the architecture's 12-hour prediction horizon ($L_{out}=12$) serves as a critical buffer against this latency. Because the model projects continuum intensity evolution 12 hours into the future, it theoretically maintains an operational visibility window of approximately 8 hours ($12 - 4$ hours processing latency), provided the emergence signature is detected within the forecast horizon. Our results indicate that the \texttt{EarlyDetect} model not only utilizes this horizon but also exhibits a raw timing bias of $-4.73$ hours (detecting signatures well before the strict onset criteria). This combination—a long prediction horizon canceling the processing delay, coupled with a hypersensitive early detection mechanism—suggests that the system remains operationally viable even under strict real-time constraints.

These results provide a strong foundation for future work. The successful architecture of \texttt{EarlyDetect} can now be applied to the more challenging but valuable task of forecasting the magnetic flux emergence, which is known to precede the continuum intensity decrease. Furthermore, this work was limited to 1D temporal sequences from individual tiles; a natural successor would be a spatio-temporal model that incorporates information from neighboring tiles, potentially providing more contextual data to improve the forecast further. Finally, the stark difference in performance between the baseline models and the Early Detection model highlights a critical takeaway for the field: for space weather forecasting, where advance warning is the primary objective, low RMSE must not be mistaken for a useful forecast. 

\section{Conclusion} \label{sec:conclusion}

In this work, we address the need for timely and accurate forecasting of solar active region emergence. We developed and evaluated a series of Transformer-based models in a systematic ablation study, comparing a novel Early Detection architecture against a standard Transformer baseline, with and without a temporal Conv1D front-end. We measured success not only by statistical accuracy (RMSE) but by predictive timing ($\Delta$ Emergence), a metric crucial for operational space weather.
We found that the standard Transformer architecture, while achieving low error, failed to produce a timely forecast, lagging the emergence by over 8 hours. The addition of a Conv1D front-end was detrimental, increasing model complexity and harming performance. Our results identify a distinct trade-off between the reactive stability of recurrent models and the proactive sensitivity of attention-based architectures. The specialized Early Detection Transformer (\texttt{EarlyDetect}), which paired a biased attention mechanism with a timing-aware loss function, was the only configuration to provide a true advance forecast, predicting the emergence 4.73 hours early. It also achieved the best statistical accuracy, with an overall RMSE of 0.1189 and an Emergence RMSE of 0.1450, representing a 10.6\% improvement over the LSTM baseline. Crucially, these results were established under a stricter, uniform emergence criterion ($k=4$ hours) compared to previous studies, ensuring that the reported advance warnings represent sustained physical emergence rather than transient noise.
However, our analysis suggests that specific magnetic morphologies may present stability challenges that challenge the Transformer's attention mechanism more than the conservative gates of the LSTM. Future work must expand the evaluation to a larger cohort of Active Regions to smooth out these potential selection biases and definitively establish the stability of the attention-based approach.

The observed performance characteristics of our Transformer-based model suggest several directions for architectural refinement. While the attention mechanism excels at capturing global temporal patterns, the quadratic computational complexity may limit scalability to longer sequences. Recent state-space models~\citep{gu2024mamba, pandey2024comparative} offer linear-time alternatives that maintain competitive performance while enabling efficient processing of extended temporal contexts. Moreover, models implementing hyperbolic geometry~\citep{patil2025hyperbolic} leverage spaces with constant negative curvature to efficiently embed hierarchical structures, demonstrating superior performance on tasks requiring multi-scale reasoning as compared to their Euclidean counterparts. Such architectural diversity can provide a rich landscape for future AR emergence prediction systems, balancing computational efficiency, interpretability, and representational capacity.

This study demonstrates that Transformer architectures, when specifically modified to prioritize early-onset signals, are highly effective for AR emergence prediction. Our findings prove that simply applying deep learning is insufficient; a generic model will optimize for a statistical fit, not for an early warning. Critically, our ablation study reveals that removing Conv1D smoothing (pure \texttt{EarlyDetect}) is preferable for warning systems because the cost of false positives (noise) is outweighed by the benefit of significantly earlier detection. The specialized, timing-aware architecture provides a new, robust baseline for future research in heliophysics and a clear path toward building operationally viable forecasting systems that prioritize advance warning over statistical smoothness.

\section{Code and Data Availability}

The code, pre-trained models, and evaluation scripts used to produce the results presented in this manuscript are publicly available. The software is distributed under the MIT license and can be accessed at the GitHub repository: \url{https://github.com/jonastirona/ar-emergence-transformers}.

The repository includes:
\begin{itemize}
    \item Pre-trained model checkpoints for all five evaluated configurations (LSTM baseline and four Transformer variants)
    \item Complete training and evaluation pipelines implemented in PyTorch
    \item Model complexity profiling tools
    \item Comprehensive documentation, including data format specifications and reproducibility instructions
    \item SLURM job scripts for HPC environments
\end{itemize}

The software requires Python 3.10+ and PyTorch 2.1.0 or later. Complete installation instructions and dependencies are provided in the repository's \texttt{README.md} file. The data used in this study can be obtained from the SolARED portal (\url{https://sun.njit.edu/sarportal/}).

\acknowledgments
This work was supported by the New Jersey Institute of Technology (NJIT) Grace Hopper AI Research Institute (GHAIRI) Seed Grant (Grant No. 179025). We thank the computing resources provided by the High Performance Computing (HPC) facility at NJIT. This work is supported by the NASA AI/ML HECC Expansion Program, and the NASA grants 23-HGIO23\textunderscore2-0077, 20-HSR20\textunderscore2-0037, 80NSSC19K0630, 80NSSC19K0268, 80NSSC20K1870, and 80NSSC22M0162. The authors acknowledge the use of ChatGPT, Gemini, and Cursor for assistance with language editing and coding.

\section*{Conflict of Interest}
The authors declare no conflicts of interest relevant to this study.

%
%

\bibliography{agusample}

\newpage
\appendix

\section{Hyperparameter Configuration}\label{app:hyperparams}

To ensure reproducibility, we provide the complete hyperparameter search spaces and optimal configurations for all models evaluated in this study. Tables \ref{tab:hyperparams_baseline} through \ref{tab:hyperparams_earlydetect_conv1d} present both the search ranges explored and the optimal values selected for each model.

\begin{table}[h]
\centering
\caption{Hyperparameter optimization for the Baseline (LSTM) Model. The optimal configuration (128-dimension embedding, 3 encoder layers) was selected from this search space, resulting in one of the smaller models tested.}
\label{tab:hyperparams_baseline}
\footnotesize
\begin{tabular}{@{}lll@{}}
\toprule
\textbf{Parameter} & \textbf{Search Range} & \textbf{Optimal Value} \\
\midrule
Input Sequence Length ($L_{in}$) & Fixed & 110 \\
Output Horizon ($L_{out}$) & Fixed & 12 \\
Embedding Dimension ($d_{model}$) & \{128, 256\} & 128 \\
Attention Heads & \{4\} & 4 \\
Encoder Layers & \{3, 5\} & 3 \\
Dropout & \{0.0, 0.3\} & 0.0 \\
Learning Rate & \{1$\times$10$^{-4}$, 1$\times$10$^{-3}$\} & 1$\times$10$^{-3}$ \\
Temporal Conv1D & \{False\} & False \\
\bottomrule
\end{tabular}
\end{table}

\begin{table}[h]
\centering
\caption{Hyperparameter optimization for the \texttt{Baseline+Conv1D} Model. The optimal configuration required a larger embedding (256-dim) and more layers (5) than the standard baseline, indicating a more complex model was chosen during the search.}
\label{tab:hyperparams_baseline_conv1d}
\footnotesize
\begin{tabular}{@{}lll@{}}
\toprule
\textbf{Parameter} & \textbf{Search Range} & \textbf{Optimal Value} \\
\midrule
Input Sequence Length ($L_{in}$) & Fixed & 110 \\
Output Horizon ($L_{out}$) & Fixed & 12 \\
Embedding Dimension ($d_{model}$) & \{128, 256\} & 256 \\
Attention Heads & \{4\} & 4 \\
Encoder Layers & \{3, 5\} & 5 \\
Dropout & \{0.0, 0.3\} & 0.0 \\
Learning Rate & \{1$\times$10$^{-4}$, 1$\times$10$^{-3}$\} & 1$\times$10$^{-4}$ \\
Temporal Conv1D & \{True\} & True \\
\bottomrule
\end{tabular}
\end{table}

\begin{table}[h]
\centering
\caption{Hyperparameter optimization for the \texttt{EarlyDetect} Model (Best Performer). This table includes the search space for the specialized loss weights. The optimal configuration favored a strong penalty on late predictions ($\lambda_{timing}=0.4$) and a moderate derivative penalty ($\lambda_{early}=0.3$), alongside a 6-layer, 8-head architecture.}
\label{tab:hyperparams_earlydetect}
\footnotesize
\begin{tabular}{@{}lll@{}}
\toprule
\textbf{Parameter} & \textbf{Search Range} & \textbf{Optimal Value} \\
\midrule
Input Sequence Length ($L_{in}$) & Fixed & 110 \\
Output Horizon ($L_{out}$) & Fixed & 12 \\
Embedding Dimension ($d_{model}$) & \{256, 512\} & 256 \\
Attention Heads & \{4, 8\} & 8 \\
Encoder Layers & \{4, 6\} & 6 \\
Dropout & \{0.0, 0.1\} & 0.0 \\
Learning Rate & \{5$\times$10$^{-5}$, 1$\times$10$^{-4}$, 2$\times$10$^{-4}$\} & 2$\times$10$^{-4}$ \\
Temporal Conv1D & \{False\} & False \\
\midrule
\textit{Early Detection Parameters} & & \\
$\lambda_{timing}$ & \{0.2, 0.3, 0.4\} & 0.4 \\
$\lambda_{early}$ & \{0.1, 0.2, 0.3\} & 0.3 \\
Timing Bias Weight & \{0.1, 0.2, 0.3, 0.5\} & 0.1 \\
Early Detection Weight ($w_{early}$) & \{0.1, 0.2, 0.3\} & 0.3 \\
Internal Threshold ($\delta$) & \{-0.003, -0.005, -0.007\} & -0.007 \\
\bottomrule
\end{tabular}
\end{table}

\begin{table}[H]
\centering
\caption{Hyperparameter optimization for the \texttt{EarlyDetect+Conv1D} Model. This model, which had the highest computational cost, settled on the largest embedding dimension (512) but a smaller number of layers (4). The optimal loss weights ($\lambda_{timing}=0.2, \lambda_{early}=0.2$) were weaker than the more successful non-Conv1D variant.}
\label{tab:hyperparams_earlydetect_conv1d}
\footnotesize
\begin{tabular}{@{}lll@{}}
\toprule
\textbf{Parameter} & \textbf{Search Range} & \textbf{Optimal Value} \\
\midrule
Input Sequence Length ($L_{in}$) & Fixed & 110 \\
Output Horizon ($L_{out}$) & Fixed & 12 \\
Embedding Dimension ($d_{model}$) & \{256, 512\} & 512 \\
Attention Heads & \{4, 8\} & 4 \\
Encoder Layers & \{4, 6\} & 4 \\
Dropout & \{0.0, 0.1\} & 0.0 \\
Learning Rate & \{5$\times$10$^{-5}$, 1$\times$10$^{-4}$, 2$\times$10$^{-4}$\} & 5$\times$10$^{-5}$ \\
Temporal Conv1D & \{True\} & True \\
\midrule
\textit{Early Detection Parameters} & & \\
$\lambda_{timing}$ & \{0.2, 0.3, 0.4\} & 0.2 \\
$\lambda_{early}$ & \{0.1, 0.2, 0.3\} & 0.2 \\
Timing Bias Weight & \{0.1, 0.2, 0.3, 0.5\} & 0.3 \\
Early Detection Weight ($w_{early}$) & \{0.1, 0.2, 0.3\} & 0.1 \\
Internal Threshold ($\delta$) & \{-0.003, -0.005, -0.007\} & -0.007 \\
\bottomrule
\end{tabular}
\end{table}

\begin{table}[H]
\centering
\caption{Configuration for the LSTM Baseline Model \citep{kasapis2025prediction} Note, in this model, the hidden size (64) is significantly smaller compared to the Transformer embedding dimensions.}
\label{tab:best_config_lstm}
\footnotesize
\begin{tabular}{@{}lr@{}}
\toprule
\textbf{Hyperparameter} & \textbf{Value} \\
\midrule
Input Sequence Length ($L_{in}$) & 110 \\
Output Horizon ($L_{out}$) & 12 \\
Hidden Size & 64 \\
Number of Layers & 3 \\
Learning Rate & 0.01 \\
Training Epochs & 1000 \\
\bottomrule
\end{tabular}
\end{table}

\section{Performance of models for test active regions}\label{app:per_ar_results}

Tables \ref{tab:ar11698_metrics} through \ref{tab:ar13183_metrics} present comprehensive performance metrics for all models across five test active regions. 

\begin{table}[h]
\centering
\caption{Detailed metrics for AR 11698. This region illustrates the high-variance nature of the \texttt{EarlyDetect} model. While it achieves an early mean timing difference of -9.75h (compared to the LSTM's +23.00h lag), the extreme standard deviation ($\pm$50.11h) reflects its volatility: it detects emergence very early on specific tiles (e.g., Tile 51) but lacks the spatial consistency of the baseline models.}
\label{tab:ar11698_metrics}
\footnotesize
\begin{tabular}{@{}lcccc@{}}
\toprule
\textbf{Model} & \textbf{Overall RMSE} & \textbf{Emergence RMSE} & \textbf{Emergence R$^2$} & \textbf{Timing $\Delta$ (h)} \\
\midrule
LSTM Baseline & 649.48 $\pm$ 167.86 & 736.86 $\pm$ 284.69 & -8.42 $\pm$ 11.05 & +23.00 $\pm$ 21.99 \\
\texttt{Baseline} & 712.97 $\pm$ 252.38 & 796.05 $\pm$ 299.52 & -20.39 $\pm$ 32.29 & -15.50 $\pm$ 59.50 \\
\texttt{Baseline+Conv1D} & 710.99 $\pm$ 248.03 & 895.66 $\pm$ 326.68 & -13.46 $\pm$ 15.55 & +24.00 $\pm$ 21.97 \\
\texttt{EarlyDetect} & 659.36 $\pm$ 236.06 & 756.02 $\pm$ 298.05 & -12.37 $\pm$ 19.05 & -9.75 $\pm$ 50.11 \\
\texttt{EarlyDetect+Conv1D} & 690.66 $\pm$ 232.17 & 975.47 $\pm$ 344.66 & -14.64 $\pm$ 18.49 & +26.00 $\pm$ 21.39 \\
\bottomrule
\end{tabular}
\end{table}

\begin{table}[H]
\centering
\caption{Detailed metrics for AR 11726. This region represents a success case for early detection across all architectures. The \texttt{EarlyDetect} model achieves the earliest mean timing (-9.40h), slightly outperforming the LSTM (-8.60h). However, the high standard deviation in the Transformer's Emergence RMSE indicates that while it captures the timing effectively, the magnitude of the predicted drop varies significantly across tiles compared to the recurrent baseline.}
\label{tab:ar11726_metrics}
\footnotesize
\begin{tabular}{@{}lcccc@{}}
\toprule
\textbf{Model} & \textbf{Overall RMSE} & \textbf{Emergence RMSE} & \textbf{Emergence R$^2$} & \textbf{Timing $\Delta$ (h)} \\
\midrule
LSTM Baseline & 2291.49 $\pm$ 1763.68 & 2971.70 $\pm$ 2163.67 & -36.24 $\pm$ 41.60 & -8.60 $\pm$ 16.30 \\
\texttt{Baseline} & 1348.95 $\pm$ 1066.91 & 1330.51 $\pm$ 384.99 & -1.82 $\pm$ 2.59 & -6.20 $\pm$ 13.18 \\
\texttt{Baseline+Conv1D} & 1547.44 $\pm$ 1262.10 & 1165.09 $\pm$ 340.77 & -0.91 $\pm$ 1.51 & -7.40 $\pm$ 15.70 \\
\texttt{EarlyDetect} & 1357.54 $\pm$ 956.57 & 1459.25 $\pm$ 420.69 & -3.86 $\pm$ 5.79 & -9.40 $\pm$ 18.01 \\
\texttt{EarlyDetect+Conv1D} & 1620.11 $\pm$ 1463.07 & 1370.81 $\pm$ 91.56 & -4.44 $\pm$ 7.82 & -10.50 $\pm$ 13.16 \\
\bottomrule
\end{tabular}
\end{table}

\begin{table}[h]
\centering
\caption{Performance metrics for AR 13165. This region provides the strongest evidence for the \texttt{EarlyDetect} architecture's sensitivity. It achieved a substantial mean advance warning of -26.25 hours, drastically outperforming the Baseline (+10.5h late) and the LSTM (-20.75h). The large standard deviation in timing ($\pm$31.18h) reflects the model's distinct response to precursors in specific tiles (e.g., Tile 31) versus others.}
\label{tab:ar13165_metrics}
\footnotesize
\begin{tabular}{@{}lcccc@{}}
\toprule
\textbf{Model} & \textbf{Overall RMSE} & \textbf{Emergence RMSE} & \textbf{Emergence R$^2$} & \textbf{Timing $\Delta$ (h)} \\
\midrule
LSTM Baseline & 658.94 $\pm$ 359.44 & 620.47 $\pm$ 290.33 & -4.20 $\pm$ 5.78 & -20.75 $\pm$ 18.61 \\
\texttt{Baseline} & 413.24 $\pm$ 140.97 & 506.84 $\pm$ 179.88 & -1.39 $\pm$ 1.04 & +10.50 $\pm$ 28.46 \\
\texttt{Baseline+Conv1D} & 512.54 $\pm$ 138.48 & 448.83 $\pm$ 156.79 & -1.10 $\pm$ 1.41 & -6.00 $\pm$ 19.61 \\
\texttt{EarlyDetect} & 454.67 $\pm$ 153.67 & 623.33 $\pm$ 245.98 & -2.62 $\pm$ 1.72 & -26.25 $\pm$ 31.18 \\
\texttt{EarlyDetect+Conv1D} & 438.86 $\pm$ 190.73 & 404.64 $\pm$ 94.73 & -0.74 $\pm$ 1.07 & -4.00 $\pm$ 20.40 \\
\bottomrule
\end{tabular}
\end{table}

\begin{table}[h]
\centering
\caption{Performance metrics for AR 13179. This region highlights a failure mode for the attention-based models in the presence of shallow or ambiguous signatures. All Transformer variants predicted the emergence significantly late (e.g., +17.00h for \texttt{EarlyDetect}), whereas the LSTM baseline proved more robust, achieving synchronous detection with a timing difference of +0.67h.}
\label{tab:ar13179_metrics}
\footnotesize
\begin{tabular}{@{}lcccc@{}}
\toprule
\textbf{Model} & \textbf{Overall RMSE} & \textbf{Emergence RMSE} & \textbf{Emergence R$^2$} & \textbf{Timing $\Delta$ (h)} \\
\midrule
LSTM Baseline & 844.94 $\pm$ 454.76 & 726.72 $\pm$ 193.19 & -1.92 $\pm$ 1.34 & +0.67 $\pm$ 7.36 \\
\texttt{Baseline} & 825.32 $\pm$ 734.06 & 876.58 $\pm$ 323.44 & -2.74 $\pm$ 1.51 & +21.00 $\pm$ 4.24 \\
\texttt{Baseline+Conv1D} & 850.56 $\pm$ 712.22 & 706.83 $\pm$ 281.18 & -1.31 $\pm$ 0.74 & +13.00 $\pm$ 2.00 \\
\texttt{EarlyDetect} & 781.12 $\pm$ 818.34 & 994.92 $\pm$ 372.22 & -3.53 $\pm$ 1.19 & +17.00 $\pm$ 0.50 \\
\texttt{EarlyDetect+Conv1D} & 781.12 $\pm$ 698.30 & 802.18 $\pm$ 366.18 & -1.54 $\pm$ 0.62 & +17.00 $\pm$ 2.94 \\
\bottomrule
\end{tabular}
\end{table}

\begin{table}[H]
\centering
\caption{Performance metrics for AR 13183. Characterized by a smooth intensity profile, this region was generally detected late by all models. However, \texttt{EarlyDetect} yielded the best performance of the group, achieving the closest timing (+4.75h) and the lowest Overall RMSE (328.27). The Conv1D front-end proved detrimental here, worsening timing to +8.00h and increasing error.}
\label{tab:ar13183_metrics}
\footnotesize
\begin{tabular}{@{}lcccc@{}}
\toprule
\textbf{Model} & \textbf{Overall RMSE} & \textbf{Emergence RMSE} & \textbf{Emergence R$^2$} & \textbf{Timing $\Delta$ (h)} \\
\midrule
LSTM Baseline & 416.41 $\pm$ 268.03 & 516.80 $\pm$ 276.63 & -1.05 $\pm$ 1.96 & +7.50 $\pm$ 4.61 \\
\texttt{Baseline} & 328.78 $\pm$ 165.90 & 667.83 $\pm$ 44.04 & -3.64 $\pm$ 5.01 & +8.75 $\pm$ 3.56 \\
\texttt{Baseline+Conv1D} & 410.38 $\pm$ 222.10 & 807.40 $\pm$ 252.02 & -3.27 $\pm$ 3.24 & +11.33 $\pm$ 1.25 \\
\texttt{EarlyDetect} & 328.27 $\pm$ 143.74 & 516.93 $\pm$ 146.41 & -2.54 $\pm$ 4.86 & +4.75 $\pm$ 2.86 \\
\texttt{EarlyDetect+Conv1D} & 389.59 $\pm$ 175.88 & 706.81 $\pm$ 209.34 & -3.01 $\pm$ 4.18 & +8.00 $\pm$ 4.24 \\
\bottomrule
\end{tabular}
\end{table}

{
\setlength{\tabcolsep}{3pt} 
\small 
\begin{longtable}{l|cccccccc}
\caption{Detailed emergence predictions by tile. Values indicate the lead time (in hours) relative to emergence onset. Negative values (e.g., -26h) indicate successful early detection (Good), while positive values indicate late detection (Bad). This granular view highlights the stability-sensitivity trade-off: while the \texttt{EarlyDetect} model secures the best global mean, it exhibits higher variance in specific regions (e.g., AR 11698) compared to the more conservative LSTM baseline.} \label{tab:detailed_predictions_long} \\
\toprule
\textbf{Model} & \textbf{T1} & \textbf{T2} & \textbf{T3} & \textbf{T4} & \textbf{T5} & \textbf{T6} & \textbf{T7} & \textbf{Overall} \\
\midrule
\endfirsthead
\multicolumn{9}{c}%
{{\bfseries \tablename\ \thetable{} -- continued from previous page}} \\
\toprule
\textbf{Model} & \textbf{T1} & \textbf{T2} & \textbf{T3} & \textbf{T4} & \textbf{T5} & \textbf{T6} & \textbf{T7} & \textbf{Overall} \\
\midrule
\endhead
\hline \multicolumn{9}{r}{{Continued on next page}} \\ \hline
\endfoot
\bottomrule
\endlastfoot

\multicolumn{9}{c}{\textbf{AR 11698 (Tiles 47--53)}} \\
\makecell[l]{Baseline+C1D} & \makecell{FP} & FN & 10h & 62h & 13h & 11h & Quiet & 24h \\
Baseline & Quiet & FN & -40h & 63h & 12h & -97h & Quiet & -15h \\
EarlyDetect & Quiet & FN & -41h & 61h & -70h & 11h & \makecell{FP} & -9h \\
\makecell[l]{EarlyDetect+C1D} & \makecell{FP} & FN & 12h & 63h & 15h & 14h & Quiet & 26h \\
LSTM & Quiet & FN & 8h & 61h & 12h & 11h & Quiet & 23h \\
\midrule

\multicolumn{9}{c}{\textbf{AR 11726 (Tiles 38--44)}} \\
\makecell[l]{Baseline+C1D} & Quiet & Quiet & 12h & -27h & 3h & -25h & 0h & -7h \\
Baseline & Quiet & Quiet & 1h & -25h & 7h & -19h & 5h & -6h \\
EarlyDetect & Quiet & Quiet & 18h & -31h & 0h & -27h & -7h & -9h \\
\makecell[l]{EarlyDetect+C1D} & Quiet & Quiet & FN & -26h & 4h & -21h & 1h & -10h \\
LSTM & Quiet & Quiet & 14h & -27h & 3h & -27h & -6h & -8h \\
\midrule

\multicolumn{9}{c}{\textbf{AR 13165 (Tiles 29--35)}} \\
\makecell[l]{Baseline+C1D} & Quiet & Quiet & -33h & 2h & 13h & FN & Quiet & -6h \\
Baseline & Quiet & Quiet & -31h & 49h & 8h & 16h & Quiet & 10h \\
EarlyDetect & Quiet & Quiet & -67h & 2h & 6h & -46h & \makecell{FP} & -26h \\
\makecell[l]{EarlyDetect+C1D} & Quiet & Quiet & -32h & 16h & 4h & FN & Quiet & -4h \\
LSTM & Quiet & Quiet & -33h & -12h & 5h & -43h & Quiet & -20h \\
\midrule

\multicolumn{9}{c}{\textbf{AR 13179 (Tiles 38--44)}} \\
\makecell[l]{Baseline+C1D} & Quiet & Quiet & FN & 15h & 11h & Quiet & Quiet & 13h \\
Baseline & Quiet & Quiet & 27h & 18h & 18h & \makecell{FP} & Quiet & 21h \\
EarlyDetect & Quiet & Quiet & 21h & 16h & 14h & Quiet & Quiet & 17h \\
\makecell[l]{EarlyDetect+C1D} & Quiet & Quiet & FN & 17h & 16h & Quiet & Quiet & 16h \\
LSTM & Quiet & Quiet & -8h & 0h & 10h & \makecell{FP} & Quiet & 0h \\
\midrule

\multicolumn{9}{c}{\textbf{AR 13183 (Tiles 38--44)}} \\
\makecell[l]{Baseline+C1D} & Quiet & Quiet & FN & 11h & 13h & 10h & Quiet & 11h \\
Baseline & Quiet & Quiet & 14h & 9h & 8h & 4h & Quiet & 8h \\
EarlyDetect & Quiet & Quiet & 9h & 5h & 4h & 1h & \makecell{FP} & 4h \\
\makecell[l]{EarlyDetect+C1D} & Quiet & Quiet & 14h & 8h & 8h & 2h & Quiet & 8h \\
LSTM & Quiet & Quiet & 14h & 7h & 8h & 1h & Quiet & 7h \\
\end{longtable}
}

\section{Continuum intensity forecast for the test active regions}\label{app:plots}

\begin{figure}[H]
\centering
\includegraphics[width=1\textwidth]{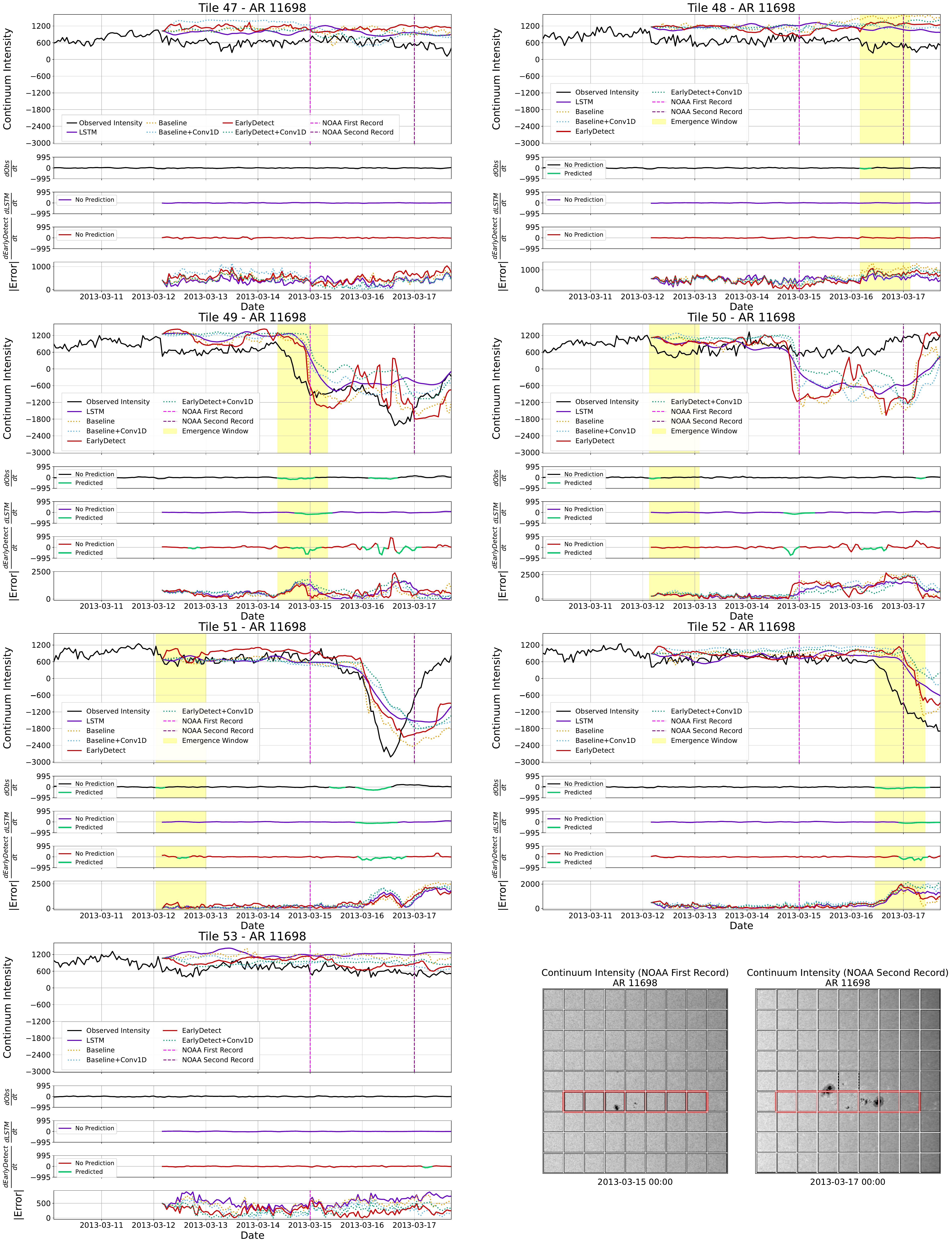}
\caption{Comparison plot for AR 11698. This figure illustrates the high-variance behavior reported in Table \ref{tab:ar11698_metrics}. While the \texttt{EarlyDetect} model achieves a mean timing of -9.75h (early) on this active region, the visual forecast is volatile, characterized by sharp pre-emergence drops in specific tiles (e.g., Tile 49). The LSTM, while smoother, lacks this aggressive early sensitivity.}
\label{fig:ar_11698}
\end{figure}

\begin{figure}[h]
\centering
\includegraphics[width=1\textwidth]{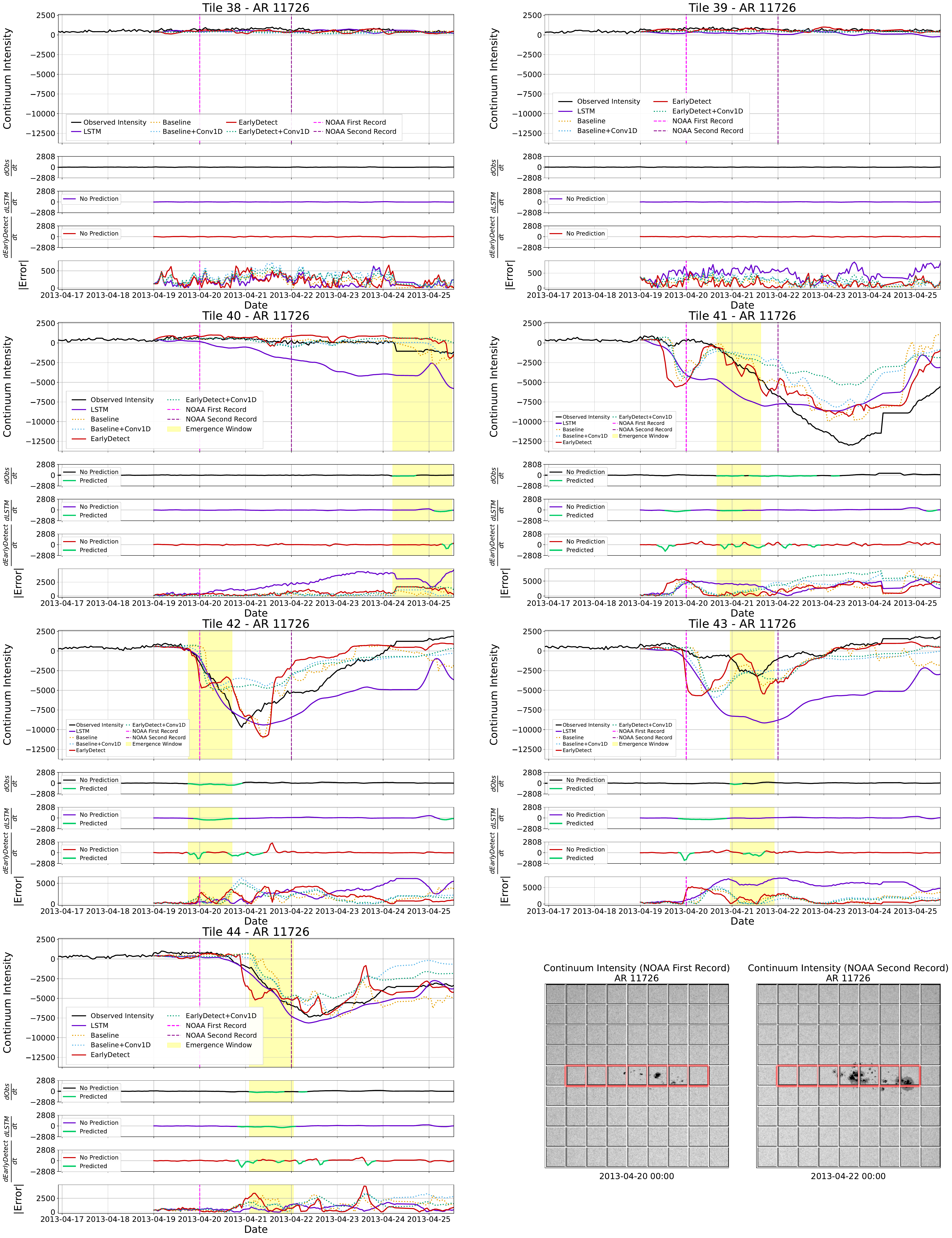}
\caption{Comparison plot for AR 11726. This region represents a consistent success case for the Transformer architecture. The \texttt{EarlyDetect} predictions (red line) are visibly shifted to the left of the observed intensity drop (black line) across the active tiles, confirming the quantitative finding of a -9.40 hour mean advance warning. Note: This AR does have missing values in the data towards the end of the selected timeline, shown by the straight line from 4-24 to 4-25.}
\label{fig:ar_11726}
\end{figure}

\begin{figure}[h]
\centering
\includegraphics[width=1\textwidth]{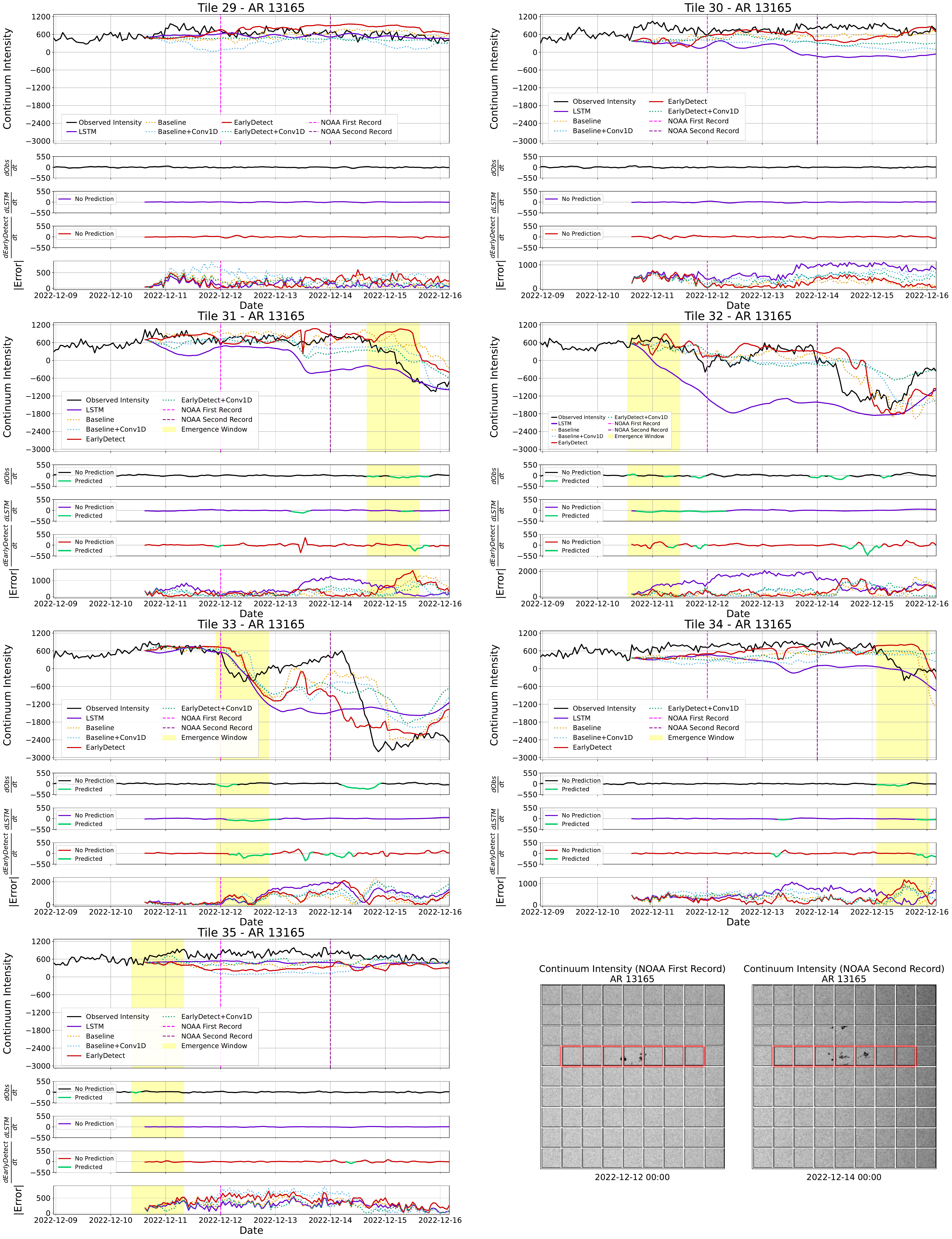}
\caption{Comparison plot for AR 13165. This region exhibits the most significant advance warning capability of the \texttt{EarlyDetect} model. The characteristic ``early dip'' is clearly visible, particularly in Tile 31, where the model anticipates the emergence well in advance. This aggressive sensitivity drives the -26.25 hour mean timing difference reported in Table \ref{tab:ar13165_metrics}.}
\label{fig:ar_13165}
\end{figure}

\begin{figure}[h]
\centering
\includegraphics[width=1\textwidth]{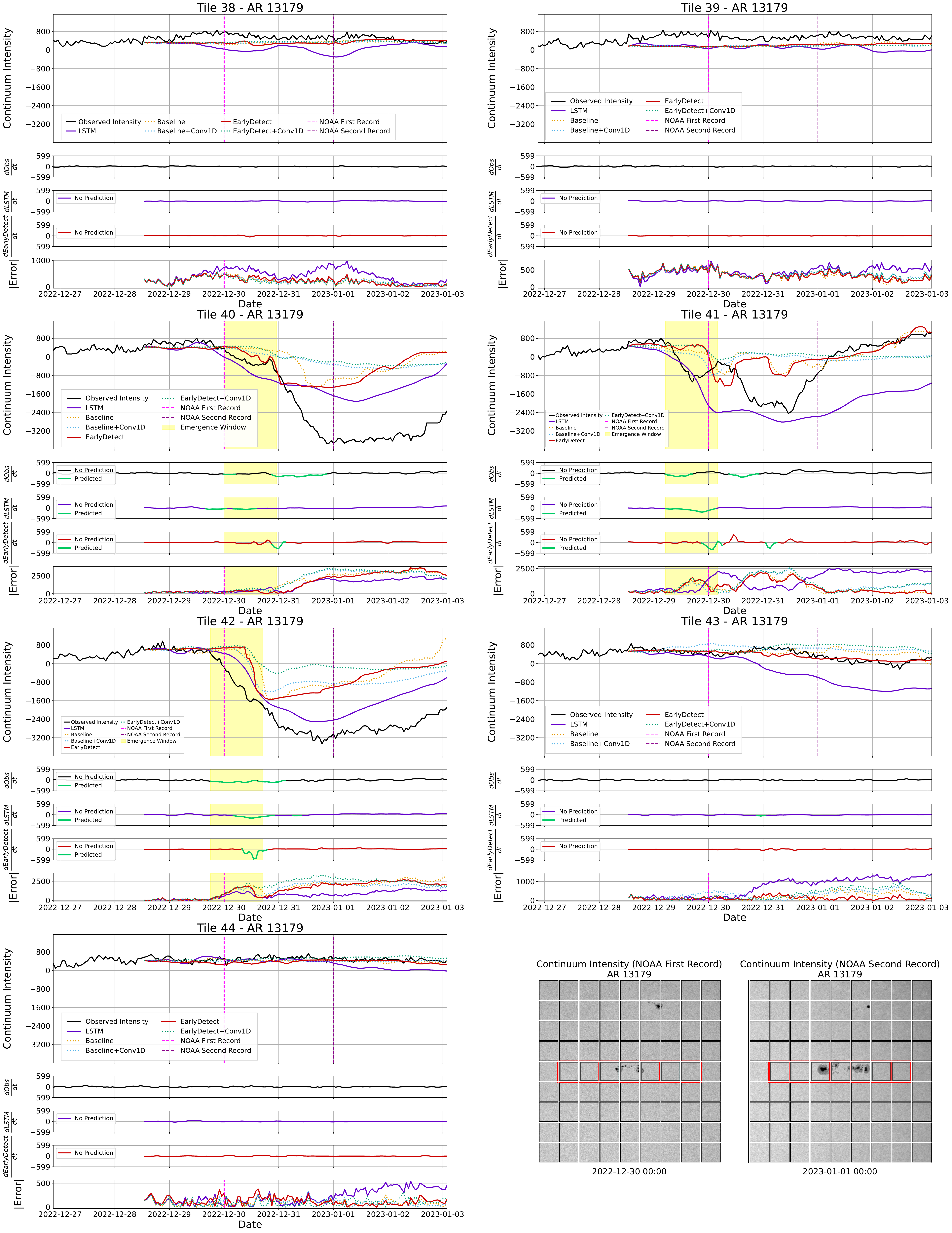}
\caption{Comparison plot for AR 13179. This region represents a failure case for the attention-based models in the presence of shallow emergence signatures. The Transformer forecasts are visibly late, trailing the event by +17 hours, whereas the LSTM baseline (purple) maintains a closer alignment (+0.67h) to the observed emergence onset.}
\label{fig:ar_13179}
\end{figure}

Figures \ref{fig:ar_11698} through \ref{fig:ar_13179} illustrate the multi-model forecasts for the test active regions. These plots follow the same format as Figure \ref{fig:all_models_comparison_AR13183} in the main text. Note: AR 11726 contains a period of missing data, appearing as a flat line in the forecast plots.

\end{document}